\documentclass[aps,prc,floatfix,showpacs,twocolumn,nofootinbib]{revtex4-1}
\usepackage{ulem}
\usepackage{dcolumn}
\usepackage{bm}
\usepackage{color}
\usepackage{amssymb}
\usepackage{amsmath}
\usepackage{graphicx}
\usepackage{amsfonts}
\usepackage{slashed}
\usepackage{pstricks}
\usepackage{float}
\usepackage{hyperref}
\usepackage{array}
\allowdisplaybreaks
\usepackage{dcolumn}
\usepackage{epsf}
\usepackage{float}
\usepackage{afterpage}
\usepackage{tikz}
\usepackage[style=base]{caption}
\usepackage[style=base]{subcaption}
\usepackage{multirow}
\usepackage{dcolumn} % for aligning tables based on decimals
\newcolumntype{d}[1]{D{.}{.}{#1}} 
\newcommand\Tstrut{\rule{0pt}{2.6ex}} % = Top strut for tables

\newcommand{\Train}{\mathbb{T}} % generic training set (really a set of indices)
\newcommand{\Test}{\mathbb{O}} % generic testing/'out of sample' set (really a set of indices)
\newcommand{\Valid}{\mathbb{V}} % generic validation set (really a set of indices)

\usepackage[ruled,vlined]{algorithm2e}

\newcommand{\vect}[1]{\boldsymbol{#1}}

\begin{document}
\title{Machine learning-based inversion of nuclear responses}
\author{
{Krishnan} Raghavan$^{\, {\rm a} }$,
%ORCID: 0000-0001-9409-2011
{Prasanna} Balaprakash$^{\, {\rm a} }$,
%ORCID: 0000-0002-0292-5715
{Alessandro} Lovato$^{\, {\rm b,c,d} }$,
{Noemi} Rocco$^{\, {\rm b,e} }$,
{Stefan M.} Wild$^{\, {\rm a,f} }$
%ORCID: 0000-0002-6099-2772
}
\affiliation{
$^{\,{\rm a}}$\mbox{Mathematics and Computer Science Division, Argonne National Laboratory, Lemont, Illinois 60439, USA}\\
$^{\,{\rm b}}$\mbox{Physics Division, Argonne National Laboratory, Lemont, Illinois 60439, USA}\\
$^{\,{\rm c}}$\mbox{Computational Science Division, Argonne National Laboratory, Lemont, Illinois 60439, USA}\\
$^{\,{\rm d}}$\mbox{INFN-TIFPA Trento Institute of Fundamental Physics and Applications, Via Sommarive, 14, 38123 {Trento}, Italy}\\
$^{\,{\rm e}}$\mbox{Theoretical Physics Department, Fermi National Accelerator Laboratory, P.O.\ Box 500, Batavia, Illinois 60510, USA}
$^{\,{\rm f}}$\mbox{NAISE, Northwestern University, Evanston, Illinois 60208, USA}
}
\date{\today}
%%%%
\begin{abstract} 
A microscopic description of the interaction of atomic nuclei with external electroweak probes is required for elucidating aspects of short-range nuclear dynamics and for the correct interpretation of neutrino oscillation experiments. Nuclear quantum Monte Carlo methods infer the nuclear electroweak response functions from their Laplace transforms. Inverting the Laplace transform is a notoriously ill-posed problem; and Bayesian techniques, such as maximum entropy, are typically used to reconstruct the original response functions in the quasielastic region. In this work, we present a physics-informed artificial neural network architecture suitable for approximating the inverse of the Laplace transform. Utilizing simulated, albeit realistic, electromagnetic response functions, we show that this physics-informed artificial neural network outperforms maximum entropy in both the low-energy transfer and the quasielastic regions, thereby allowing for robust calculations of electron scattering and neutrino scattering on nuclei and inclusive muon capture rates.
\end{abstract}
%\pacs{24.10.Cn,25.30.Pt,26.60.-c}
%%%%%
\maketitle

\section{Introduction}
Electron scattering experiments are powerful tools to simultaneously investigate the short- and long-range many-body dynamics of atomic nuclei. These experiments contributed to demonstrating the limitations of an independent particle picture of the nucleus that fails to provide a fully quantitative description of atomic nuclei~\cite{Benhar:2006wy}. At large momentum transfer, the large excess of neutron-proton correlated pairs with respect to the proton-proton and neutron-neutron pairs has highlighted the importance of the tensor component of the nuclear interaction and the interplay between nucleonic and partonic degrees of freedom~\cite{Wiringa:2013ala,Atti:2015eda,Hen:2016kwk}. The field has experienced a renewed interest also in view of its interplay with high-precision measurements of neutrinos and their oscillations~\cite{Amaro:2019zos}. This is the main focus of the accelerator-based neutrino oscillation program, which includes ongoing experiments such as NOvA~\cite{nova_web} and T2K~\cite{t2k_web} and planned ones such as DUNE~\cite{dune_web} and Hyper-K~\cite{hk_web}. Nuclear targets are utilized in the detectors to increase the event rate. Hence, the determination of oscillation parameters requires accurate theoretical calculations of neutrino-nucleus interactions in a broad range of energy, in which a variety of reaction mechanisms are at play~\cite{Benhar:2015wva, Katori:2016yel, Alvarez-Ruso:2017oui}. We also note that neutrino experiments utilizing the Liquid Argon Time Projection chamber technology have reached a degree of sophistication suitable to identifying short-range correlated pairs of nucleons~\cite{Acciarri:2014gev}.

In the low-energy regime, the inclusive lepton-nucleus cross section is dominated by coherent scattering, excitations of low-lying nuclear states, and collective modes. At energies on the order of hundreds of MeV, the leading mechanism is quasielastic (QE) scattering, in which the probe interacts primarily with individual nucleons bound inside the nucleus. Corrections to this leading mechanism arise from processes in which the lepton couples to interacting nucleons, via nuclear correlations and two-body currents.  

The inclusive lepton-nucleus scattering cross section is completely determined by the electroweak response functions, which hold all information about the dynamics of the nuclear target. The Green's function Monte Carlo (GFMC) method~\cite{Carlson:2014vla} has been successfully employed to compute the electromagnetic, neutral-current, and charged-current response functions of $^4$He and $^{12}$C in the QE region, up to moderate values of the momentum transfer~\cite{Carlson:2001mp,Lovato:2016gkq,Lovato:2017cux,Lovato:2020kba} and the muon capture rates of $^{4}$He and $^{3}$H~\cite{Lovato:2019fiw}. These calculations have unambiguously demonstrated the importance of properly treating nuclear correlations and meson exchange currents even for QE kinematics. Within this approach, the electroweak response functions are inferred from their Laplace transforms, denoted as Euclidean responses, that are estimated during the GFMC imaginary time propagation. Retrieving the energy dependence of the response functions from their Euclidean counterparts is nontrivial. 

The maximum entropy method (MaxEnt)~\cite{Bryan:1990,Jarrell:1996} has been extensively employed to retrieve the energy dependence of the electroweak response functions. Despite its success in the QE region, MaxEnt appears to be inadequate to precisely reconstruct the low-energy structure of the nuclear response functions. In Ref.~\cite{Lovato:2016gkq}, experimental inputs on the low-lying nuclear transitions have been utilized to properly describe the longitudinal electromagnetic responses of $^{12}$C in the low-energy region. A comparison between GFMC and exact Faddeev results for the $^{3}$H muon capture rate has contributed to exposing the shortcomings of MaxEnt in reconstructing the charged-current response functions near the nuclear breakup threshold, corresponding to energies of few MeVs~\cite{Lovato:2019fiw}. In addition, although heuristics have been used, to the best of our knowledge there is no rigorous way to propagate the statistical uncertainties of the Euclidean response into the response function and to quantify the systematic errors due to the approximate inversion of the Laplace transform. These errors would propagate into the GFMC estimates of lepton-nucleus cross sections and are critical for informative comparisons with experiments.

In recent years, an increase in available computing resources has been accompanied by a prodigious rise of techniques based on machine learning (ML), which are now ubiquitous in physics~\cite{Carleo:2019}.  Within low-energy nuclear physics, artificial neural networks (ANNs) have been used to estimate ground state energies and radii of nuclei by using results from no-core shell model calculations \cite{Negoita2019,Jiang2019}.
Gaussian process emulators were used in Ref. \cite{Neufcourt2019} for Bayesian model mixing in order to predict bound nuclides between silicon and titanium. The authors of  Ref.~\cite{Keeble:2019bkv} represent the deuteron's wave function with ANNs.  In Ref.~\cite{Adams:2020aax} ANNs were used to model the Jastrow correlator of $A\leq 4$ nuclei. Several works have demonstrated that ML approaches are suitable for solving inverse problems~\cite{McCann:2017,Arsenault:2017}. In particular, Refs.~\cite{Yoon:2018,Fournier:2020} utilized ANNs to recover the electron single-particle spectral density in the real frequency domain from the fermionic Green's function in the imaginary time domain. The same problem was tackled in Ref.~\cite{Xie:2019} by utilizing an Adams-Bashforth residual ANN. In both cases, the ANN approaches have been found to outperform MaxEnt implementations. 

In this work we develop a novel ANN architecture suitable for approximately inverting the Laplace transform of realistic nuclear electromagnetic response functions, similar to those computed with the GFMC method. The simulated responses utilized in the training dataset exhibit a sharp Gaussian peak corresponding to the low-energy elastic transition and an asymmetric broad peak in the QE region. The positions, heights, and widths of these two peaks are modeled consistently with their energy and momentum transfer behavior as measured by electron-scattering experiments. 
In contrast to previous approaches,
we incorporate physics-grounded constraints into the neural-network architecture and use an entropic cost function. We demonstrate an improved accuracy of the inversion in the low $\omega$ region with increased robustness to noise as compared with MaxEnt techniques. 
This robustness is especially relevant in view of 
%future 
applications of nuclear quantum Monte Carlo methods to the calculations of the electroweak response functions of larger nuclei relevant to the neutrino-oscillation program, including $^{16}$O and $^{40}$Ar. One such approach, the auxiliary field diffusion Monte Carlo~\cite{Schmidt:1999lik}, suffers from a more severe sign problem than the GFMC; this will in turn result in noisier Euclidean response functions.

This work is organized as follows. In Sec.~\ref{sec:responses} we state the problem to be solved and discuss the relevant features of the nuclear electromagnetic responses. In Sec.~\ref{sec:ML} we describe our ML algorithm. In Sec.~\ref{sec:res} we present our results, and in Sec.~\ref{sec:conc} we discuss our conclusions. 

\section{Nuclear Responses}
\label{sec:responses}
In the one-photon exchange approximation, the inclusive electron-nucleus scattering cross section can be expressed in terms of the longitudinal and transverse response functions, $R_L({\bf q},\omega)$ and $R_T({\bf q},\omega)$, respectively, where ${\bf q}$ and $\omega$ are the electron momentum and energy transfers. The response functions encode all information on nuclear structure and dynamics and are defined as
\begin{equation}
\begin{array}{rll}
R_\alpha ({\bf q},\omega)&= \displaystyle \sum_f & \left\langle 0 | j_\alpha^\dagger({\bf q},\omega) |f\right\rangle \left\langle f | j_\alpha({\bf q},\omega) |0\right\rangle  
\\
& & \times \delta(E_f-\omega-E_0), 
\end{array}
\label{eq:res_def}
\end{equation}
for $\alpha=L,T$. In Eq.~\eqref{eq:res_def}, $|0\rangle$ and $|f\rangle$ represent the initial and final nuclear states of energies $E_0$ and $E_f$, respectively, and $j_L({\bf q},\omega)$ and $j_T({\bf q},\omega)$ are the electromagnetic charge and current operators, respectively.

A direct calculation of $R_\alpha({\bf q},\omega)$ requires evaluating all of the individual transition amplitudes induced by the charge and current operators and is therefore impractical except for very light nuclear systems~\cite{Shen:2012xz,Golak:2018qya}. The use of integral transform techniques has proven helpful in circumventing these difficulties. One such approach is based on the calculation of the Euclidean response~\cite{Carlson:1992ga}, which corresponds to the Laplace transform
\begin{equation}
E_\alpha({\bf q},\tau) = \int_0^\infty d\omega\, e^{-\omega\tau} R_\alpha({\bf q},\omega) \, .
\label{eq:euc_def}
\end{equation}
Fixing the intrinsic energy dependence of the charge and current operators to the QE peak, $\omega_{\rm QE}=\sqrt{{\bf q}^2+m^2}-m$, where $m$ denotes the mass of the nucleon, one can express the Euclidean responses as ground-state expectation values
\begin{equation*}
E_\alpha({\bf q},\tau)=\langle 0 | j_\alpha^\dagger ({\bf q},\omega_{\rm QE}) e^{-(H-E_0)\tau} j({\bf q},\omega_{\rm QE}) | 0\rangle,
\end{equation*}
where $H$ is the nuclear Hamiltonian. These expectation values can be evaluated by using the GFMC method on a uniform grid of $n_\tau$ imaginary-time points~\cite{Carlson:1992ga,Carlson:2001mp}. A set of noisy estimates for $E_\alpha({\bf q},\tau_i)$ can be obtained by performing independent imaginary-time propagations, from which the average Euclidean response $\bar{E}_\alpha({\bf q},\tau_i)$ and the covariance $C_{ij}$ between the data at $\tau=\tau_i$ and $\tau=\tau_j$ can be readily estimated~\cite{Lovato:2016gkq}. Note that, in general, the covariance matrix $C$ is nondiagonal because of correlations among the imaginary-time points.

\subsection*{Problem statement and the MaxEnt approach} 
%GAIL - I added the asterisk so that you don't have a section A without a section B
%SW: Good.
In addition to the imaginary time $\mathcal{T}=[\tau_1, \cdots, \tau_{n_\tau}]$, we discretize the continuous variables $\omega$ on $n_\omega$ grid points and thus define $\Omega = [\omega_1, \cdots, \omega_{n_\omega}]$ and the kernel $K(\omega_i,\tau_j) =e^{-\omega_i\tau_j}\Delta\omega_i$,
where $\Delta\omega_i$ is the discretization width at $\omega_i$. Dropping, for simplicity, the momentum transfer dependence and the subscript $\alpha$ of the response functions, we can rewrite the Laplace transform of Eq.~\eqref{eq:euc_def}  in the compact matrix form
\begin{equation*}
E(\mathcal{T}) = K( \Omega, \mathcal{T}) R(\Omega),
\end{equation*}
where $E(\mathcal{T}) \in \mathbb{R}^{n_\tau}$, $R(\Omega)  \in \mathbb{R}^{n_\omega}$, and $K( \Omega, \mathcal{T}) \in \mathbb{R}^{n_\tau \times n_\omega}$. The response function can thus be formally (for an appropriate definition of $\cdot^{-1}$) recovered by
\begin{equation}
R(\Omega) = K( \Omega, \mathcal{T})^{-1} E(\mathcal{T}).
\label{eq:matrix}
\end{equation}
However, the inversion of $K(\Omega, \mathcal{T})$ is numerically unstable because of the exponentially small tails in the kernel function for large $\omega$. Retrieving the response function from noisy GFMC estimates of $E(\tau)$ involves significant difficulty; widely different response functions can correspond to very similar Euclidean responses.

Several algorithms have been developed for approximately inverting the Laplace transform by using prior knowledge about the solution. Arguably the most robust and popular of these is MaxEnt~\cite{Bryan:1990,Jarrell:1996}, which has been used to reconstruct the (smooth) energy dependence of the nuclear response functions around the QE peak~\cite{Lovato:2016gkq,Lovato:2017cux,Lovato:2020kba}. Within MaxEnt, the solution of the inverse problem is the response function that maximizes the posterior probability $P(R|E)$ (i.e., the conditional probability of $R(\Omega)$ given $E(\mathcal{T})$). Bayes' theorem states that the posterior probability is proportional to the product $P(E|R) \times P(R)$, where $P(E|R)$ is the likelihood function and $P(R)$ is the prior probability, containing information about the response function to be reconstructed. Arguments based on the central limit theorem show that the asymptotic limit of the likelihood function is given by $P(E|R) \propto \exp(-\chi^2/2)$, where 
%SW? Is a ref (textbook) helpful here?
\begin{equation}
\chi^2 = \sum_{i,j=1}^{n_\tau} \left(E(\tau_i)-\bar{E}(\tau_i)\right)C_{ij}^{-1} \left(E(\tau_j)-\bar{E}(\tau_j)\right)\, .
\label{eq:chi2}
\end{equation}

Since the response functions are positive and normalizable, they can be interpreted as  probability distributions. The principle of maximum entropy states that the values of a probability distribution are to be assigned by maximizing the entropy, which is defined by
\begin{equation}
S = \sum_{i=1}^{n_\omega} \left( R(\omega_i) - M(\omega_i) - R(\omega_i)\ln\left(\frac{R(\omega_i)}{M(\omega_i)}\right) \right)\Delta\omega_i \, . 
\label{eq_entropy}
\end{equation}
The positive-valued 
% that is: M(\omega) > 0 for all \omega
$M(\omega)$ is the default model and encodes our prior knowledge about $R(\omega)$ in the absence of data. The entropy measures how much the response function differs from the model. It vanishes when $R(\omega) = M(\omega)$ and is negative when $R(\omega)\neq M(\omega)$. 

MaxEnt  improves upon the standard $\chi^2$ minimization by using the prior information, whereby $R(\omega)$ can be interpreted as a probability distribution. For given $\bar{E}(\tau_i)$, $C_{ij}$, and default model $M(\omega_i)$, the response functions are found minimizing the quantity
\begin{equation}
Q = \frac{1}{2}\chi^2 - \alpha S\,,
\label{eq:q_def}
\end{equation}
where $\alpha$ is a fixed parameter that controls the relative importance between the entropy and the error terms. Despite its tendency to underfit the data~\cite{Titterington:1985}, in this work we adopt the  {\it historic} MaxEnt approach~\cite{Gull:1978}, which consists in choosing $\alpha$ so that $\chi^2 = n_\tau$. On the other hand, the more sophisticated {\it classic} MaxEnt~\cite{Skilling:1989} and {\it Bryan} MaxEnt~\cite{Bryan:1990}---both relying on the  probability $P(\alpha|E)$ to determine $\alpha$ -- tend to overfit the data since  $P(\alpha|E)$ is evaluated only approximately in practice~\cite{VonDerLinden:1999,Hohenadler:2005}.  
In general, the arbitrariness in choosing $\alpha$ prevents a robust reconstruction of the rich structure that characterizes the low-$\omega$ region of $R(\omega)$, without running the risk of overfitting $E(\tau)$ and hence causing spurious oscillation in the reconstructed response function. 
%\PB{not clear; in the previous paragraph, you said you are using underfitting MaxEnt to overcome the limitations of overfitting MaxNet. Also, if we know the method is underfitting, its predictive performance will be poor. In that case, why this baseline?}.
%\AL{Because MaxEnt is the "workhorse" for the inversion of the Laplace transform in condensed-matter physics. All the papers we quote on ML techniques to invert the Laplace use MaxEnt as a baseline. The problem with it is that if you use the fully Bayesian ``classic'' MaxEnt or ``Bryan'' MaxEnt you overfit the data, the ``historic'' MaxEnt instead slightly underfits data.}

\section{Physics-informed neural network}
\label{sec:ML}
As mentioned in the preceding section, the inversion of $K(\Omega, \mathcal{T})$ is numerically unstable, and retrieving $R(\Omega)$ from $E(\mathcal{T})$ is an ill-posed inverse problem. To overcome this difficulty, we seek an approximate solution by designing a physics-informed neural network, which we dub ``Phys-NN,'' that is suitable for finding a controlled approximation $\hat{R}(\Omega)$ for the right-hand side of Eq.~\eqref{eq:matrix}.
\begin{figure*}[!htb]
    \centering
    \includegraphics[width = 0.8\textwidth]{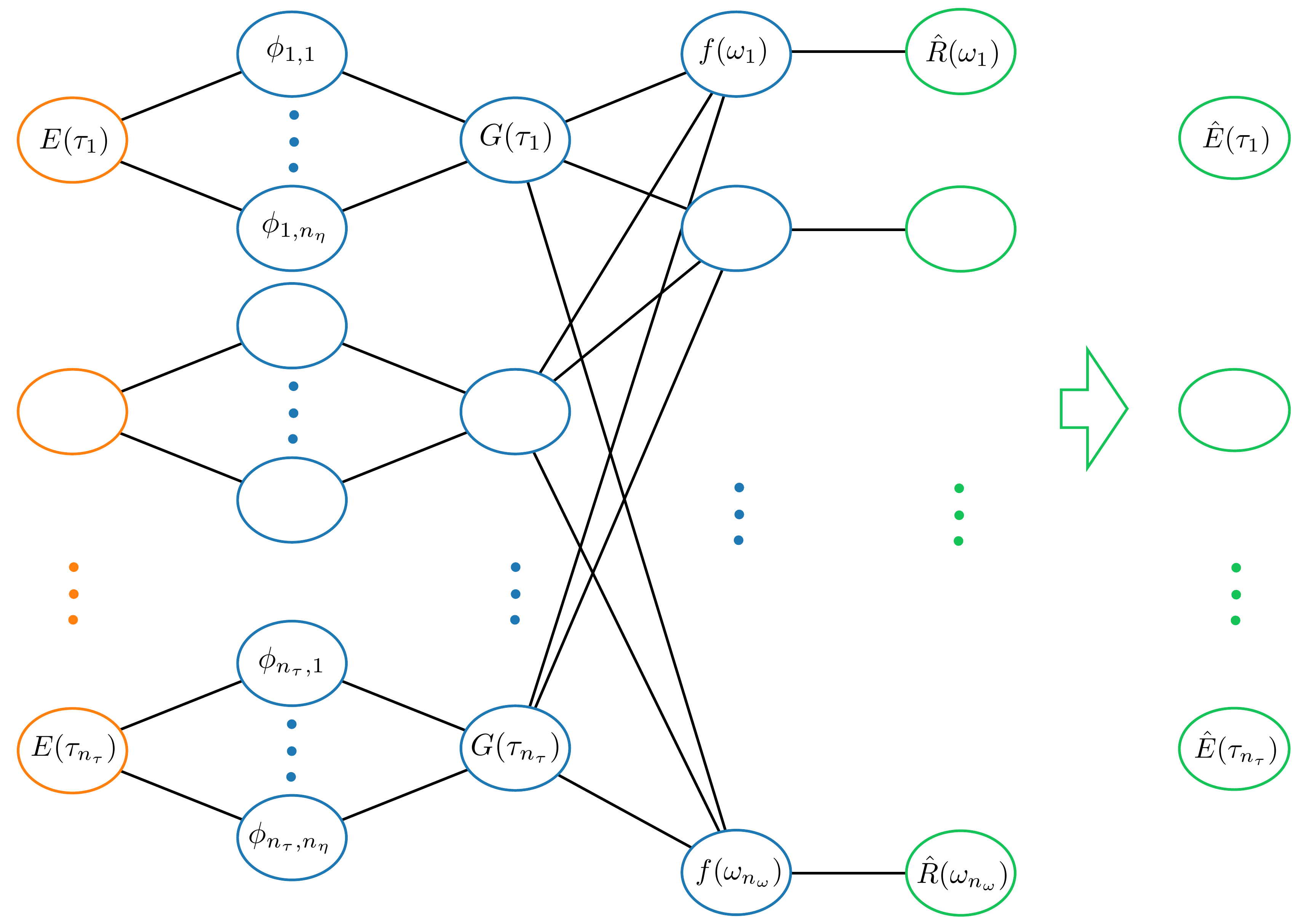}
    \caption{Schematic overview of the Phys-NN approach.}
    \label{fig:my_architecture}
\end{figure*}

\subsection{The Phys-NN model}
To model $\hat{R}(\Omega)$, we start by constructing a set $\mathcal{R}$ of basis functions that takes into account the physics of the problem, while being as broadly applicable as possible. Note that each term in the matrix $K(\Omega, \mathcal{T})$, is proportional to $e^{-\tau_j \omega_i}$ and therefore a reasonable choice to capture its structure is the Gaussian kernel basis, defined as
\begin{equation}
\label{eq:normal}
	    \phi(x,\mu,\sigma ) = \frac{1}{\sqrt{2\pi}\sigma}e^{- \frac{(x - \mu)^2}{2\sigma^2}}, \quad x \in \mathbb{R}\,.
\end{equation}
Finding the location and the scale of the Gaussian kernel, denoted by  $\mu \in \mathbb{R}$ and $\sigma>0$, respectively, is part of the ML training problem. The first layer of the neural network, whose architecture is displayed in Fig.~\ref{fig:my_architecture}, takes as input the $n_\tau$-dimensional vector $E(\mathcal{T})$. To form a basis for each 
$E(\tau)$, we apply $n_\eta$ Gaussian units of the form \eqref{eq:normal}, where $n_\eta$ is a hyperparameter.  
We then contract these Gaussian units with the resulting hidden layer outputs multiplied by weights $w_{i,j}$ to obtain the output associated with $\omega_i$. Formally, the Phys-NN is given by 
\begin{equation}
f(E(\mathcal{T}) ; \vect{\theta}) = \left[ \begin{array}{c}
	    \displaystyle\sum_{j = 1}^{n_{\tau}} W_{1,j} \sum_{k = 1}^{n_\eta}  \phi( E(\tau_{j}),\mu_{j,k}, \sigma_{j,k} ) \\
	       \vdots \\
	     \displaystyle \sum_{j = 1}^{n_{\tau}}  W_{n_{\omega},j} \sum_{k = 1}^{n_\eta} \phi( E(\tau_{j}),\mu_{j,k}, \sigma_{j,k} )
	 \end{array} \right],  
	 \label{eq_Arch}
\end{equation}
where we use $\vect{\theta}=(\vect{\mu}, \vect{\sigma},\vect{W})$ to denote the collection of training parameters 
$\vect{\mu},\vect{\sigma} \in \mathbb{R}^{n_{\tau} \times n_{\eta}}$
and $\vect{W} \in \mathbb{R}^{n_{\omega} \times n_{\tau}}$. 
We can express Eq.~\eqref{eq_Arch} componentwise by
\begin{equation*}
 f\left(\omega_i\right) = 
	    \displaystyle\sum_{j = 1}^{n_{\tau}} W_{i,j} \sum_{k = 1}^{n_\eta} \phi( E(\tau_{j}),\mu_{j,k}, \sigma_{j,k} ),
	    \; i=1,\ldots, n_{\omega}.
\end{equation*}
In order to ensure that the response function is positive for all $\omega \in \Omega$, the output is passed through an exponential function, and the final approximation of the response functions is given by
\begin{equation*}
\hat{R}(\Omega;\vect{\theta}) = \frac{1}{\mathcal{N}_0} e^{f(E(\mathcal{T}); \vect{\theta})}\, .
\end{equation*}
The normalization factor $\mathcal{N}_0$ ensures that the integral of $\hat{R}(\Omega;\vect{\theta}))$ coincides with $E(\tau_0)$, so that the output of the Phys-NN automatically satisfies the sum rule of the response function.

\subsection{Simulated data}
\begin{figure}[!b]
    \centering
    \includegraphics[width=0.9\columnwidth]{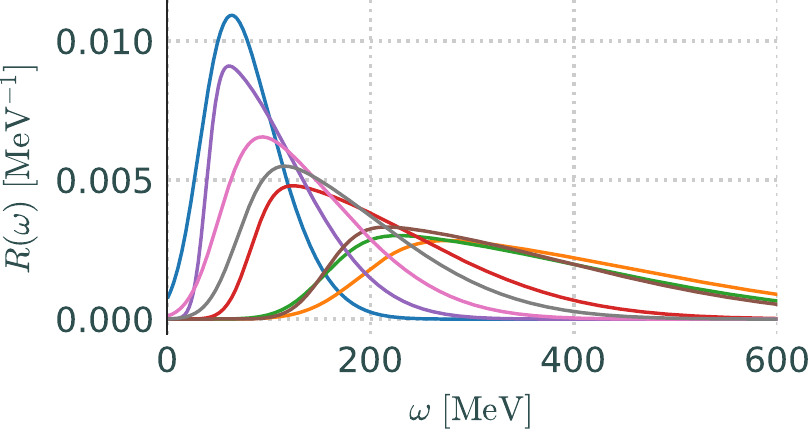}
    \caption{Training data examples of response functions exhibiting a single asymmetric QE peak.}
    \label{fig:Rw_QE}
\end{figure}

To train the Phys-NN, we use two distinct datasets of physically meaningful $R(\omega)$, $E(\tau)$ pairs that are simulated as follows. The responses belonging to the first dataset---a few of which are displayed in Fig.~\ref{fig:Rw_QE}---are characterized by a single asymmetric peak, corresponding to the QE reaction mechanism, modeled by a skew-normal distribution 
\begin{equation*}
R_{\rm QE}(\omega) = N_{\rm QE}\,\phi(\omega,\omega_{\rm QE},\sigma_{\rm QE})\Phi\left( \frac{\alpha(\omega-\omega_{\rm QE})}{\sigma_{\rm QE}}\right),
\end{equation*}
where $\phi(\omega,\omega_{\rm QE},\sigma_{\rm QE})$ is the Gaussian density defined in Eq.~\eqref{eq:normal} and 
\begin{equation*}
\Phi(x) = \frac{1}{2}\left(1 + \rm{erf}\left(\frac{x}{\sqrt{2}}\right) \right)
\end{equation*}
is the Gaussian's cumulative distribution function. The values of $N_{\rm QE}$, $\sigma_{\rm QE}$, and $\alpha$ are obtained according to arguments based on the scaling of the response functions~\cite{Donnelly:1999sw}. 

First, we sample the variable $q$, corresponding to the momentum transfer, from a uniform distribution between $100$ and $700$ MeV. Consistent with non-relativistic calculations of the electromagnetic response functions, we assume that $\omega_{\rm QE}=q^2/(2m_N)+\epsilon$, where $m_N$ is the nucleon mass and $\epsilon=25$ MeV is the nuclear binding. A suitable definition for the QE region corresponds to the interval $\psi=[-1,1]$ for the scaling variable. Hence, in the non-relativistic case, the width of the QE peak is approximately $2qk_F/m_N$, and we take the Fermi momentum to be $k_F=225$ MeV~\cite{Rocco:2017hmh}. In the simulated responses, we encode this behavior by allowing $20\%$ fluctuations of $\sigma_{\rm QE}$ around its central value $2qk_F/m_N$. The height of the quasielastic peak is $N_{\rm QE}/\sigma_{\rm QE}$ and $N_{\rm QE}$ guarantees that $R_{\rm QE}(\omega)$ is normalized to unity. The skewness parameter $\alpha$ is randomly sampled between 2 and 10---the normal distribution is recovered for $\alpha=0$. This interval has been chosen to reproduce the typical asymmetry displayed by the electromagnetic responses of light nuclei. 

\begin{figure}[t]
    \centering
    \includegraphics[width=0.9\columnwidth]{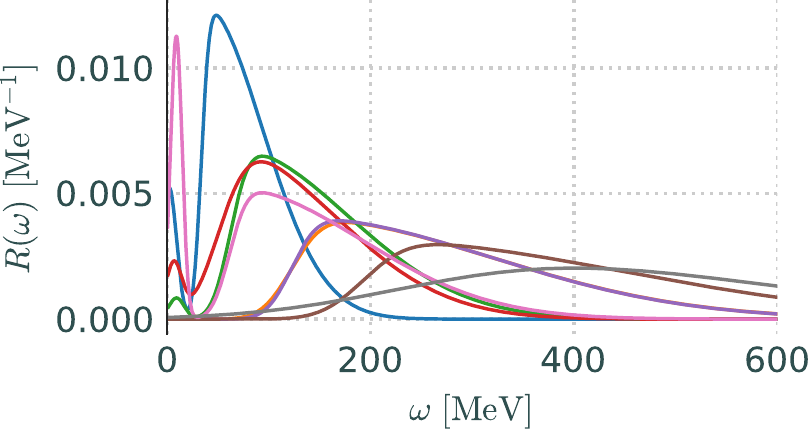}
    \caption{Training data examples of response functions characterized by an EL narrow peak in addition to the QE peak.}
    \label{fig:Rw_EL}
\end{figure}

As shown in Fig.~\ref{fig:Rw_EL}, the responses belonging to the second dataset exhibit two distinct peaks, corresponding to the elastic (EL) and QE transitions, namely, $R_{\rm EL}(\omega)+R_{\rm QE}(\omega)$.
The elastic transition contributes in the low $\omega$ region, and it is characterized by a $\delta$-like peak centered at $\omega_{\rm EL}=q^2/(2M_A)$, with $M_A\approx 4m_N$ being the mass of the $^4$He nucleus. We model the EL response with a Gaussian distribution
\begin{equation*}
    R_{\rm EL}(\omega)= \phi(\omega,\omega_{\rm EL},\sigma_{\rm EL}),
\end{equation*}
where $\sigma_{\rm EL}$ is uniformly sampled between $5$ and $10$ MeV to get a much narrower peak than the QE one. The integrated strength of the EL transition is proportional to the square of the elastic transition form factor $F_{\rm EL}(q)$. Inspired by the sum-of-Gaussians parameterizations of $F_{\rm EL}(q)$ in Ref.~\cite{Sick:2001rh}, we sample $N_{\rm EL}$ proportional to $e^{-\frac{\gamma}{2} q^2}$, where we take $\gamma=400$ MeV to reproduce the low-momentum behavior of $F_{\rm EL}(q)$ for the $^4$He nucleus. A direct consequence of this choice is that the strength of the EL peak decreases with the momentum transfer. Consistent with the one-peak case, we enforce the normalization 
\begin{equation*}
E(\tau_0) = \int d\omega \left( R_{\rm EL}(\omega)+R_{\rm QE}(\omega) \right) = 1\, .
\end{equation*}

The response functions are conveniently tabulated on a uniform $\omega$ grid between $0$ and $2$ GeV with $n_\omega=2000$. The corresponding Euclidean responses are obtained by numerically integrating $R(\omega)$. Since the simulated response are smooth functions of $\omega$, the numerical integration error on the Euclidean responses is smaller than $10^{-5}$. To mimic the statistical error of GFMC calculation, we ``corrupt'' the simulated $E(\tau)$ by adding stochastic noise~\cite{Fournier:2020}:
\begin{equation}
E(\tau_i) + \epsilon_i\, ,
\label{eq:eq_DG}
\end{equation}
where $\epsilon_i$ are independent samples from a Gaussian  distribution with mean zero and standard deviation $\sigma$. Consistent with typical GFMC calculations of the Euclidean electromagnetic responses of $^4$He, we take $\sigma=10^{-4}$ in most of our tests. 

For each of the one-peak and two-peaks cases, we generate a total of $500,000$ pairs $(R_k(\Omega),E_k(\mathcal{T}))\in \mathbb{R}^{n_\omega + n_\tau}$ of responses and corresponding Euclidean responses, which we then partition into training ($\Train$), validation ($\Valid$), and test/out-of-sample ($\Test$) datasets. The one-peak and two-peaks test datasets comprise $1,000$ pairs each; the combined test dataset is just the union of these two sets. We use 80\% and 20\% of the remaining data for training the network and validation, respectively. Since MaxEnt is relatively slow----taking about 5 seconds to perform one inversion of the Laplace transform----our comparison is limited to the test dataset.

\subsection{Training} 
Values for the parameters $\vect{\theta}$ are found by the standard supervised learning approach of approximately solving
\begin{equation}
\label{eq6}
	     \min_{\vect{\theta}}  \, \frac{1}{|\Train|} \sum_{k\in \Train} \ell\left(E_k(\mathcal{T}), R_k(\Omega), \hat{R}_k(\Omega;\vect{\theta})\right)
\end{equation}
by using a minibatch-based stochastic gradient descent procedure to minimize an empirical loss function. 
Our overall objective in Eq.~\eqref{eq6} is the average loss over the $|\Train|$ points in the training set. 
For each data and model output, we employ a loss function that is the sum of a \textit{response cost} and a \textit{Euclidean cost},
\begin{equation*} 
\ell(E_k, R_k, \hat{R}_k) = \gamma_{R} S_R(R_k, \hat{R}_k) +  \gamma_E \chi^2_{E}(E_k,\hat{R}_k),
\end{equation*}
where $\gamma_{R}, \gamma_{E} >0$ are user-defined parameters. The response cost is defined according to the entropy measure of Eq.~\eqref{eq_entropy}, namely
\begin{align}
& S_R(R,\hat{R}) = \nonumber\\
&\sum_{i=1}^{n_\omega} \left( R(\omega_i) - \hat{R}(\omega_i) - R(\omega_i)\ln\left(\frac{R(\omega_i)}{\hat{R}(\omega_i)}\right) \right)\Delta\omega_i, 
\label{eq:res_cost}
\end{align}
and ensures that the reconstructed response functions are close to the original ones. The Euclidean cost, which is aimed at aligning the Laplace transform of $\hat{R}(\Omega;\vect{\theta})$ with the original Euclidean response, is the reduced $\chi^2$ per degrees of freedom   
\begin{align}
\chi^2_{E}(E, \hat{R}) = \frac{1}{n_\tau}\sum_{j=1}^{n_\tau}  \frac{1} {\sigma_j^2} \Big(E(\tau_j)- \hat{E}(\tau_j) \Big)^{2}\, .
\label{eq:euc_cost}
\end{align}
Compared with Eq.~\eqref{eq:chi2}, in Eq.~\eqref{eq:euc_cost} we have assumed a diagonal covariance matrix, with the diagonal elements corresponding to variance of the independent Gaussian distributions of Eq.~\eqref{eq:eq_DG}: $\sigma_j^2 = \sigma^2 = 10^{-8}$ for all $j$. 
This assumption can be easily relaxed when dealing with correlated data. We evaluate $\hat{E}(\mathcal{T};\vect{\theta} ) = K(\mathcal{T}, \Omega) \hat{R}(\Omega;\vect{\theta})$ by using a simple trapezoidal rule
\begin{equation}
\hat{E}(\tau_j) = \sum_{i = 1}^{n_{\omega}} e^{-\omega_{i} \tau_{j}} \hat{R}(\omega_{i}) \Delta \omega_{i}.
\label{eq:lab}
\end{equation}
As discussed in the following section, the positive values of $\gamma_{R}$ and $\gamma_{E}$ are chosen to compensate for the fact that $\chi^{2}_{ E}(E,\hat{R})$ is typically much larger than the entropy $S_R(R,\hat{R})$.  

Since the inversion of the Laplace transform is an ill-posed problem, there are many response functions whose Laplace transform are compatible with the original Euclidean responses. Consequently, there are instances in which $\chi^2_E$ is small 
even when the reconstructed response is not similar to the original one, leading to potential instabilities in the minimization procedure. To tame this behavior, we 
split the training 
into two phases.

In the first phase, we take $\gamma_{R} = 10^7$ and $\gamma_E = 10^{-7}$ and optimize the network using the Adam~\cite{kingma2014adam} optimizer with a learning rate of $10^{-3}$. Since $\gamma_R \gg \gamma_E$, the entropy response cost dominates the loss function and drives the reconstructed response functions close to the original ones. Once the $S_R$ has reduced significantly, we enter the second phase of the optimization, where we keep $\gamma_{R} = 10^{7}$ but increase the relative importance of the Euclidean cost by taking $\gamma_E = 1$ so that Phys-NN learns to keep the Laplace transform of the response function close to the original Euclidean response. Reducing the learning rate in the second phase to $10^{-5}$ is necessary in order to keep the reconstructed response functions close to the optimal ones found in the previous phase.

The neural-network variants are implemented in Python 3.6 by using  TensorFlow 2.0 libraries~\cite{tensorflow2015-whitepaper}.  Training, validation, and testing are performed  using systems with NVIDIA Tesla V100 SXM2 GPUs with 32GB HBM2 hosted at Argonne's Joint Laboratory for System Evaluation.

\section{Results}
\label{sec:res}
We consider three  realizations of Phys-NN that differ in the datasets used for training, validation, and testing purposes: one-peak data only, two-peak data only, and combined  one-peak and two-peak data. We quantify the accuracy of our approach using three metrics averaged over the associated test/out-of-sample dataset $\Test$. We use the average entropy
$$\overline{S_R} = \frac{1}{|\Test|} \sum_{k\in \Test} S_R(R_k, \hat{R}_k),
$$
with the entropy $S_R$ defined in Eq.~\eqref{eq:res_cost}, as well as the average reduced $\chi^2_{E}$, 
$$\overline{\chi^2_{E}} = \frac{1}{|\Test|} \sum_{k\in \Test} \chi^2_{E}(E_k, \hat{R}_k),
$$
with $\chi^2_{E}$ defined in Eq.~\eqref{eq:euc_cost}. 
We also employ a metric $\overline{R_R^{2}}$ for the response functions, which is defined as an average over $|\Test|$ terms of the form  
\begin{equation}
R^2_R(R_k,\hat{R_k}) = \frac{ \sum_{i=1}^{n_\omega} (\hat{R}_k(\omega_{i}) - R_k(\omega_{i}))^2}{\sum_{i=1}^{n_\omega}(\hat{R}_k(\omega_{i}) - \bar{R}_k(\omega))^2}\, .
\end{equation}

\subsection{Out-of-sample tests} 

\begin{table}[!b]
\caption{Phys-NN and MaxEnt testing metrics $\overline{S_R}$, $1-\overline{R_R^2}$, and $\overline{\chi_E^{2}}$ for the one-peak, two-peak, and combined datasets. The standard errors on the last digit of $\overline{\chi^2_E}$ are given in parentheses.\label{tab:results2}}
\begin{tabular}{l||d{3.3}|d{2.7}|d{4.3}}
            & \multicolumn{1}{c|}{$1-\overline{R_R^{2}}$}             & \multicolumn{1}{c|}{$\overline{\chi^{2}_E}$}    & \multicolumn{1}{c}{$\overline{S_R}$}  \\[1pt]
            & \multicolumn{1}{c|}{$\times 10^{-4}$} & & \multicolumn{1}{c}{$\times 10^{-4}$} 
            \\[3pt]
            \hline \hline 
            & \multicolumn{3}{c}{\textbf{Phys-NN}} \Tstrut \\[3pt]
One-peak    & 0.42    & 1.171(13)    & 0.72  \\ 
Two-peak    & 9.04    & 3.220(87)    & 9.16    \\ 
Combined    & 0.61    & 2.335(14)    & 3.66   \\[3pt]
\hline 
            & \multicolumn{3}{c}{\textbf{MaxEnt}} \Tstrut \\[3pt]
One-peak    & 29.7 & 1.015 \; (1)     & 60.4  \\ 
Two-peak    & 84.8 & 1.016 \; (1)     & 107  \\ 
Combined    & 57.2 & 1.015 \; (1)     & 83.7  
\end{tabular}
\end{table}
The values for the three testing metrics for the single-peak, two-peak, and combined datasets are listed in Table~\ref{tab:results2}. 
For both Phys-NN and MaxEnt, the one-peak reconstructions are the closest to their original inputs, the two-peak reconstructions are the worst, and the combined dataset reconstructions rest between those of the other two datasets. This behavior is not unexpected, since the response functions characterized by two peaks, with the EL one in the low-$\omega$ region, are notoriously more difficult to reconstruct than those having a single broad QE peak.

\begin{figure*}[t]
\begin{subfigure}{0.29\textwidth}
    \includegraphics[width = \columnwidth]{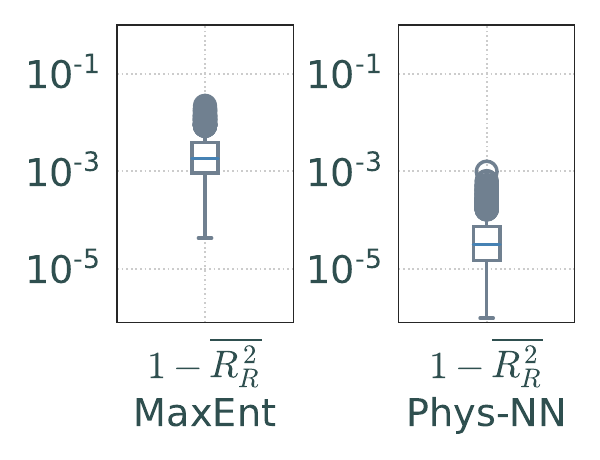}
\end{subfigure}
\begin{subfigure}{0.29\textwidth}
    \includegraphics[width = \columnwidth]{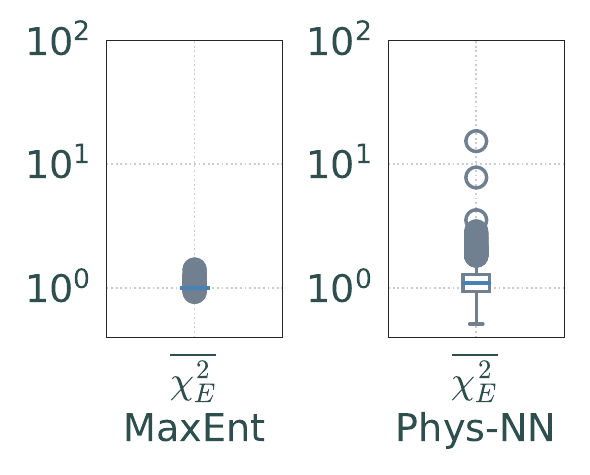}
\end{subfigure}
\begin{subfigure}{0.29\textwidth}
    \includegraphics[width = \columnwidth]{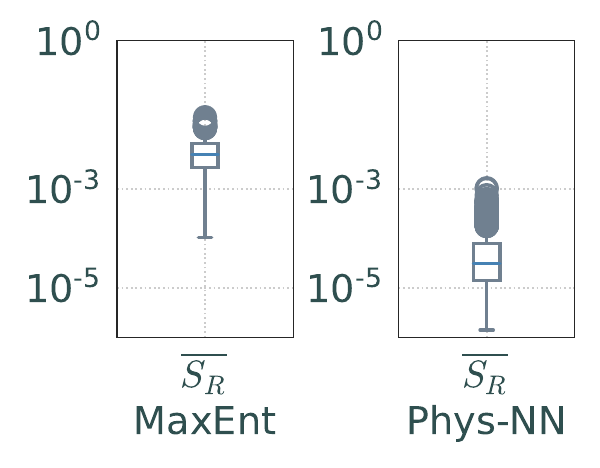}
\end{subfigure}\\
\begin{subfigure}{0.29\textwidth}
    \includegraphics[ width = \columnwidth]{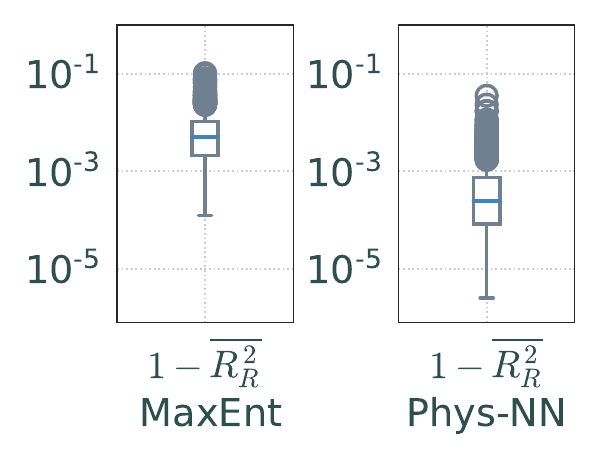}
\end{subfigure}
\begin{subfigure}{0.29\textwidth}
    \includegraphics[ width = \columnwidth]{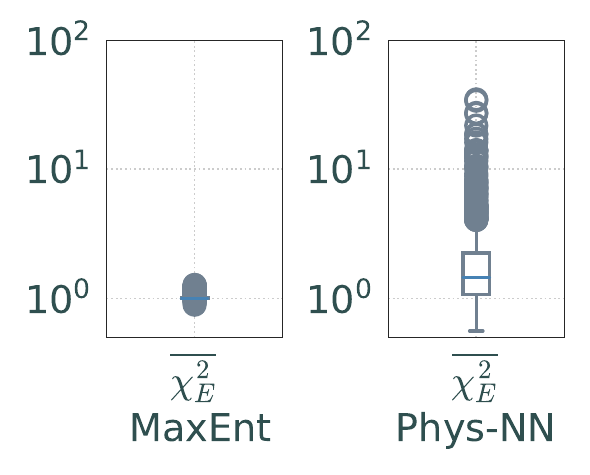}
\end{subfigure}
\begin{subfigure}{0.29\textwidth}
    \includegraphics[width = \columnwidth]{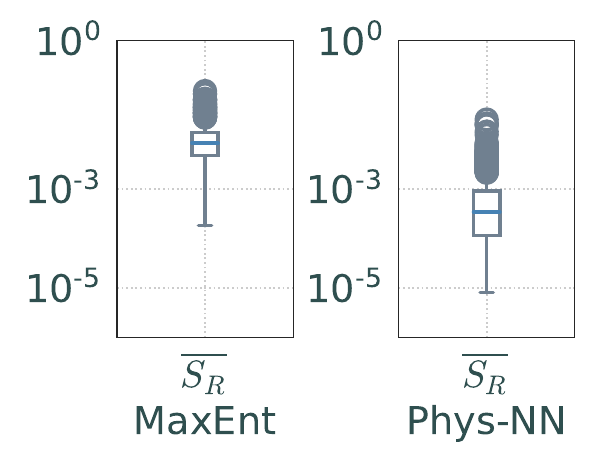}
\end{subfigure}
\caption{Box plots of (left) $R^2$, (middle) $\chi_E^2$, and (right) $S_R$ for the Phys-NN and MaxEnt methods. The top and bottom rows refer to the one-peak and two-peaks datasets, respectively. The line in the middle of the box denotes the median, and the box represents the range between the 25\% and 75\% quantiles. Whiskers cover the area between the 1\% and 99\% quantiles; data beyond these whiskers are outliers and are indicated by circles.}
\label{fig:results_box}
\end{figure*}

For Phys-NN, the one-peak response function metrics  $1-\overline{R_R^{2}}$ and $\overline{S_R}$ are on the order of $10^{-5}$. The reduced $\chi^2$ is also close to one;  smaller values indicate potential overfitting~\cite{Birge:1932}. 
When reconstructing responses belonging to the two-peak dataset, we observe slightly worse, although still satisfactory, performance 
compared with the one-peak case, as quantified by the larger values of all three metrics; for the combined dataset, 
%The accuracy of the inversion benefits from the enlarged dataset: $1-\overline{R_R^{2}}$ and $\overline{S_R}$ are both very small, indicating good agreement between reconstructed $\hat{R}(\Omega)$ with the exact $R(\Omega;\vect{\theta})$, and  
$\overline{\chi^2_E}$ is only slightly larger than 2. 

In Table~\ref{tab:results2} one can see in what ways Phys-NN  outperforms MaxEnt: both the $1-\overline{R_R^{2}}$ and $\overline{S_R}$ values obtained with MaxEnt are significantly worse, up to two orders of magnitude, than those of Phys-NN. This is a clear indication that Phys-NN  captures the energy dependence of the response functions better than does MaxEnt. Since historic MaxEnt finds the optimal response function by fixing $\alpha$ of Eq.~\eqref{eq:q_def} so that $\chi^2_E=1$, it is not surprising that MaxEnt's reduced $\chi^2$ values are closer to one than those found by Phys-NN. As evidenced by the other two metrics, because of the ill-posed nature of the problem, achieving $\chi^2_E\approx 1$ does not guarantee an accurate reconstruction of the original response functions. 

\begin{figure}[!hb]
\begin{subfigure}{0.48\linewidth}
    \includegraphics[keepaspectratio,width = \columnwidth]{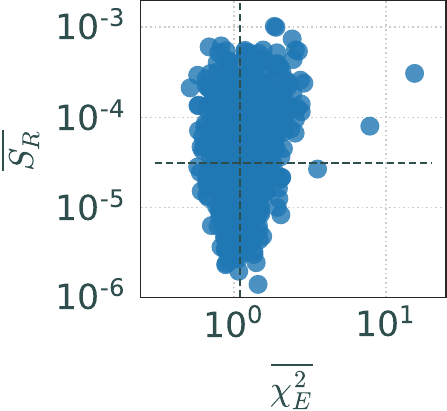}
\end{subfigure}
\begin{subfigure}{0.48\linewidth}
    \includegraphics[keepaspectratio, width = \columnwidth]{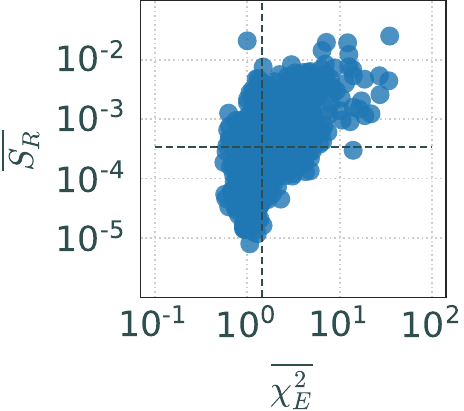}
\end{subfigure}\\
\begin{subfigure}{0.48\linewidth}
    \includegraphics[keepaspectratio, width = \columnwidth]{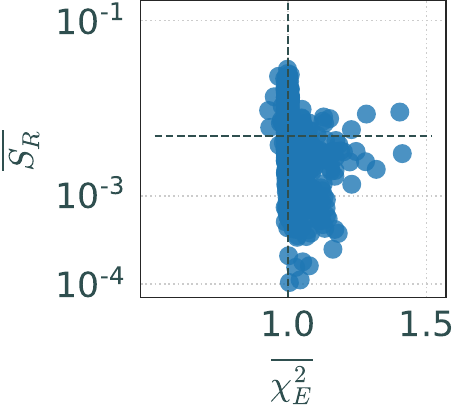}
\end{subfigure}
\begin{subfigure}{0.48\linewidth}
    \includegraphics[keepaspectratio, width = \columnwidth]{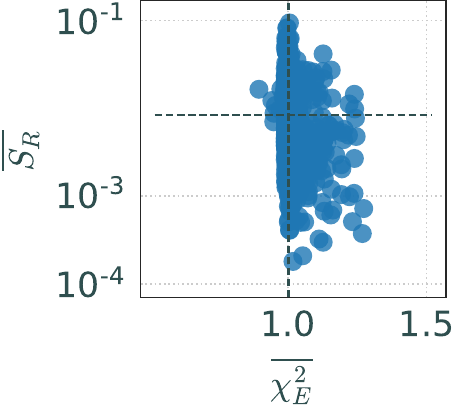}
\end{subfigure}
\caption{Correlation plots of $\overline{\chi^2_{E}}$ versus $\overline{S_R}$ as obtained with the Phy-NN (top row) and MaxEnt (bottom row) methods. The left and right columns refer to the one-peak dataset and the two-peak dataset, respectively. The reference lines indicate the median $\overline{\chi^2_{E}}$ and $\overline{S_R}$ values.}
\label{fig:corr}
\end{figure}

\begin{figure*}
\includegraphics[ width = \textwidth]{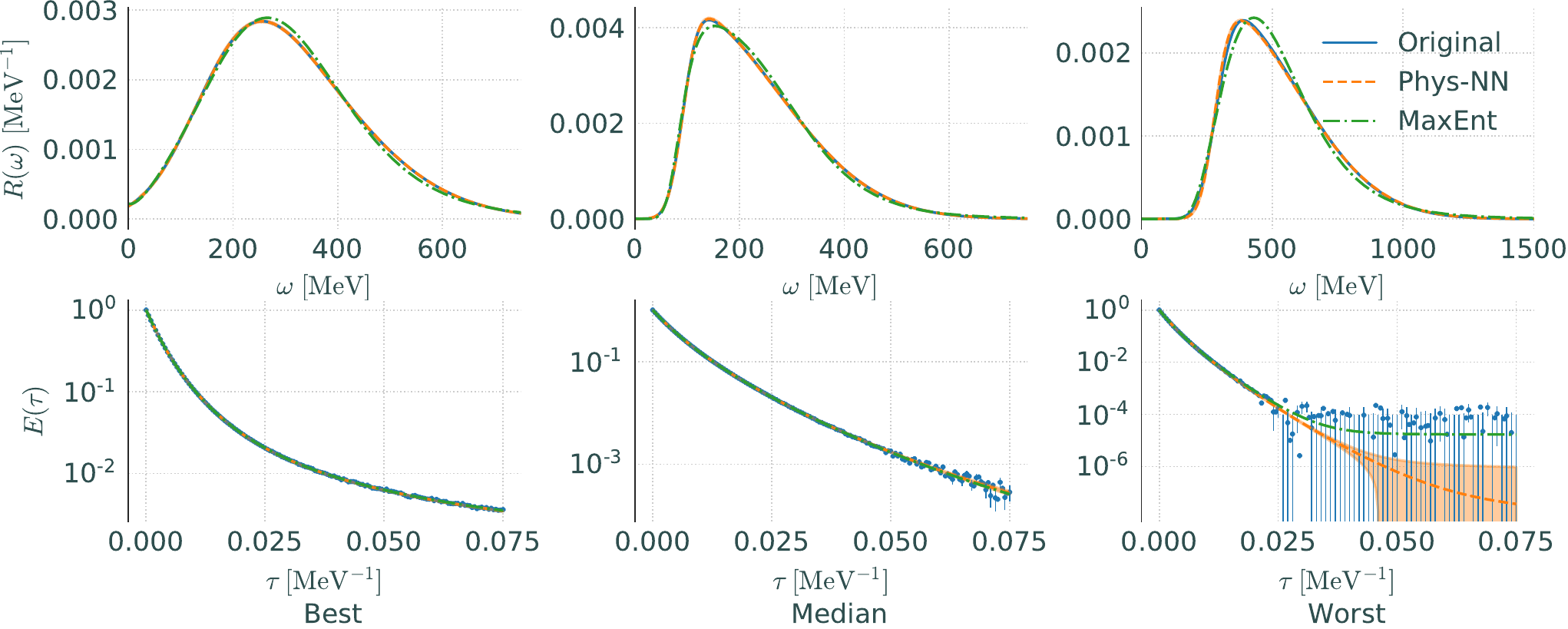}
\caption{Comparison between the Phys-NN and MaxEnt reconstructions for the one-peak dataset. The top row displays the response functions and the bottom row the corresponding Euclidean responses.}
\label{fig:results_recon_one}
\end{figure*}

\begin{figure*}
\includegraphics[width = \textwidth]{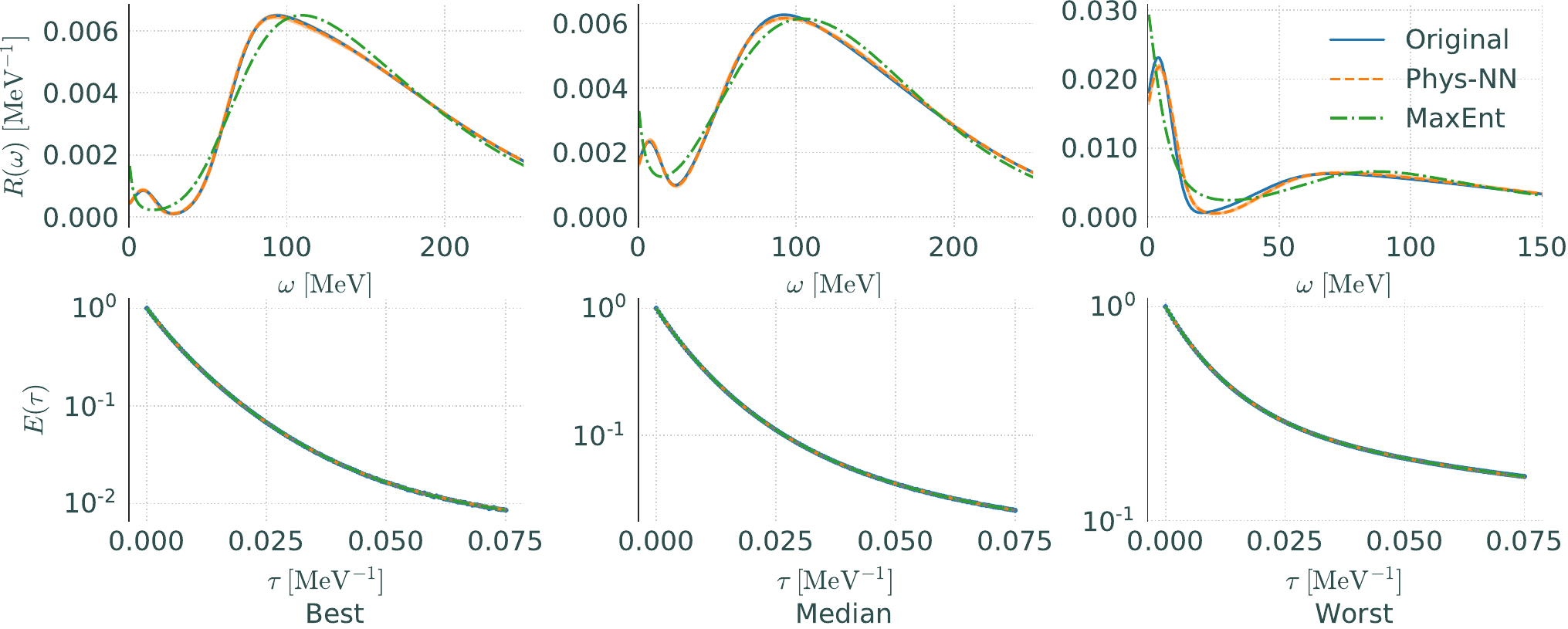}
\caption{Same as Fig.~\ref{fig:results_recon_one} for the two-peaks dataset.}
    \label{fig:results_recon_two}
\end{figure*}

To further examine the performance of Phy-NN and MaxEnt, in Fig.~\ref{fig:results_box} we display box plots of the distributions of the $1-\overline{R_R^{2}}$, $\overline{S_R}$, and $\overline{\chi_E^{2}}$ metrics for the one-peak (top row) and two-peak (bottom row) datasets. Consistent with the results listed in Table~\ref{tab:results2}, for both Phy-NN and MaxEnt, the one-peak $1-\overline{R_R^{2}}$ and $\overline{S_R}$ distributions are narrower and centered on smaller values than are the two-peak ones, while the combined dataset results are intermediate between the two. Since Phys-NN is trained to keep the reconstructed response function as close as possible to the original ones, we observe a much smaller spread of $1-\overline{R_R^{2}}$ and $\overline{S_R}$ values compared with MaxEnt. This behavior, which is exhibited across the one-peak, two-peak, and combined datasets, provides additional support for Phys-NN's reconstruction performance.

Because  the historic MaxEnt algorithm is based on $\overline{\chi^2_E}$ minimization, the resulting distributions of $\overline{\chi^2_E}$ for both the one-peak dataset and the two-peak dataset are narrow and centered on one. The spread associated with the Phys-NN results is larger. 
To investigate  correlations between $\overline{\chi^{2}_E}$ and $\overline{S_R}$, in Fig.~\ref{fig:corr} we show scatter plots for the one-peak and two-peak datasets. Some correlation is visible in the Phys-NN results, displayed in the top two panels, especially for the two-peak dataset. 
%SW: I do not think you need to make this next point; Phys-NN is trying to get good agreement in both E and R. If anything, my conclusion would be that when Phys-NN is bad in E, then it is bad in R.  
%AL: Ok... I guess what I was trying to say is that in a real-world scenario, the only indicator that we can use is chi2E. While this give little information when using MaxEnt, it could be used to determine the accuracy of Phys-NN.
%As a consequence, the difference between the reconstructed Euclidean response and the original response can be used as an indicator for the accuracy of the Phys-NN inversion procedure.
Conversely, the MaxEnt scatter plots show no correlation between $\overline{\chi^2_E}$ and $\overline{S_R}$, since the $\overline{\chi^2_E}$ values are relatively constant around one, even for widely different $\overline{S_R}$. The correlations between $\overline{\chi^2_E}$ and $1-\overline{R_R^2}$ exhibit an almost identical pattern and are thus not included here.  

Direct comparison of Phys-NN and MaxEnt outputs is presented in Fig.~\ref{fig:results_recon_one}, where we display the Phys-NN \textit{best} (left panels), \textit{average} (central panels), and \textit{worst} (right panels) reconstructed response functions, 
%SW: for this ordering, a critic would might say that it's not fair to judge MaxEnt on the best (or even median?) of the Phys-NN results. I think the ordering here is OK, but want to make sure others are aware.
%AL: yes, agreed. However, it is fair to say that the worst of Phys-NN is certainly better than the worst of MaxEnt. 
according to the $\overline{S_R}$ values of the Phys-NN results, and the corresponding Euclidean responses from the one-peak dataset. Here, the training is performed on the combined dataset, to better test whether Phys-NN is able to learn how to simultaneously reconstruct one-peak and two-peak response functions. The uncertainty associated with the random initialization of the Phys-NN parameters is estimated by performing ten independent training procedures, each corresponding to a distinct random seed used by the training procedure. We gather the predictions obtained from each of these ten runs to estimate the error band displayed by the shaded area in Fig.~\ref{fig:results_recon_one}. Not only the best and the average but also the worst response functions reconstructed with the Phys-NN are in better agreement with the original ones than are those obtained with the MaxEnt algorithm. The Laplace transform of the Phys-NN response functions are also in excellent agreement with the original Euclidean responses: the $\overline{\chi^2_E}$ values are $1.071$, $0.902$, and $1.834$ for the best, average, and worst reconstructions, respectively. As discussed previously, 
by design the MaxEnt $\overline{\chi^2_E}$ values are all very close to one. 

An analogous pattern emerges in the two-peak dataset. In this case, the best and the average Phys-NN responses, represented in the left and central of  Fig.~\ref{fig:results_recon_two}, respectively, are in excellent agreement with the original ones. Only minor discrepancies are visible in the worst reconstruction, displayed in the right panels. Although larger than in the one-peak case, the Phys-NN reduced $\overline{\chi^2_E}$ values are more than satisfactory: the values for the best, average, and worst reconstructions are $1.102$, $1.024$, and $6.996$, respectively. This behavior is reflected in the excellent agreement between the original and reconstructed Euclidean responses. On the other hand, despite the MaxEnt values for $\overline{\chi^2_E}$ again being very close to one, MaxEnt  consistently fails to resolve the EL peak in the low-energy region. In addition, it often yields QE peaks that are shifted to higher energy transfer than in the original response functions.  

\begin{figure}[b]
\centering
 \includegraphics[width = 0.49\columnwidth]{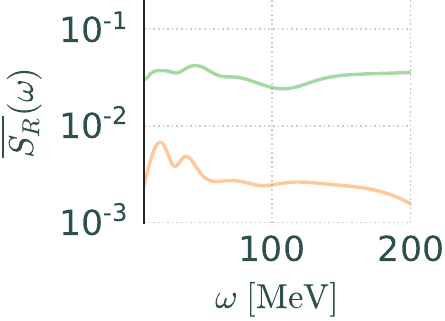}
\includegraphics[width =0.49\columnwidth]{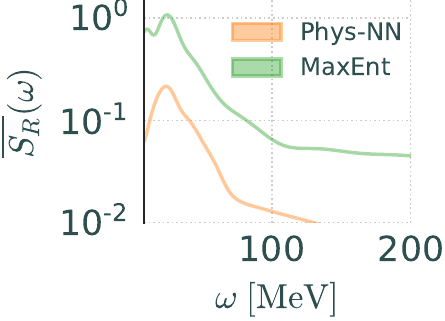}
\caption{Energy-dependent entropy for the Phys-NN and MaxEnt results for the one-peak (left panel) and two-peak (right panel) datasets.}
\label{fig:results_comparison_omega_noise}
\end{figure}

Among the shortcomings of the MaxEnt technique, the most problematic is probably its poor performance in the low-energy transfer region. The results shown in Fig.~\ref{fig:results_recon_two} clearly indicate that Phys-NN performs much better there. To quantify this behavior, we define an $\omega$-dependent entropy, $S_R(\omega)$, by restricting the integral of Eq.~\eqref{eq:res_cost} to an interval of $5$ MeV around each value of the energy transfer grid $\omega_i$ in the region $0<\omega<200$ MeV. First, we compute $S_R(\omega)$ for all the responses in the test datasets; then we calculate the average and the standard error of this quantity, displayed by the shaded areas in Fig.~\ref{fig:results_comparison_omega_noise} for the one-peak (left panel) and two-peak (right panel) case. The Phys-NN results are consistently below the MaxEnt ones, indicating better reconstruction performance for both one-peak and two-peak data. This fact will likely have important implications for GFMC calculations of the inclusive lepton-nucleus cross section in the low-energy regime. 

\begin{figure}[b]
    \centering
    \includegraphics[width = 0.88\columnwidth]{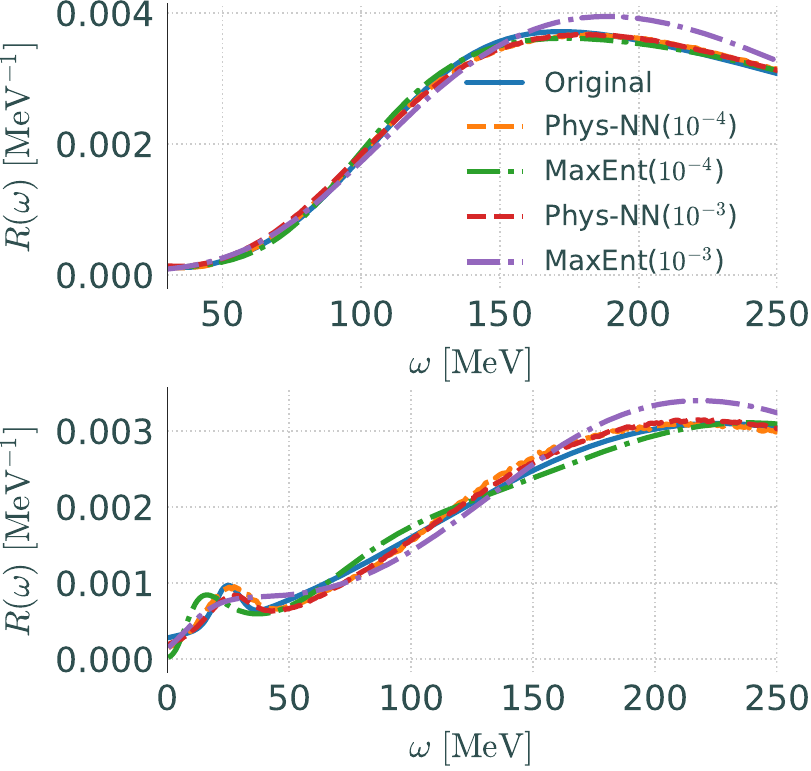}
\caption{Phys-NN and MaxEnt reconstruction performance with increasing level of noise in the input Euclidean responses.}
\label{fig:results_noise}
\end{figure}

\subsection{Predictions on noisier inputs} 
An important feature of any reconstruction technique is its robustness to the noise level of the input Euclidean response functions. We analyze how the performance of the Phys-NN and MaxEnt methods deteriorate when the standard deviation of the Gaussian noise of Eq.~\eqref{eq:eq_DG} is increased from $\sigma=10^{-4}$ to $\sigma=10^{-3}$. 
%To investigate this behavior, we train and test Phys-NN using the combined dataset with a noise level of $\sigma=10^{-3}$. 
For the results in this section, we indicate the dataset used for training by including the training data standard deviation in parentheses.
Training is always done on the combined dataset, and the training strategy and hyperparameters are unchanged from those used for the noise level $\sigma = 10^{-4}$. 

In Fig.~\ref{fig:results_noise} we compare sample reconstructed response functions when the noise on the input Euclidean is increased from $\sigma = 10^{-4}$ to $\sigma = 10^{-3}$. In both the one-peak (top panel) and two-peak (bottom panel) response,  MaxEnt clearly is more susceptible to the increased noise level than is Phys-NN. In the one-peak case, MaxEnt($10^{-3}$) significantly overestimates the height of the QE peak and shifts its maximum to higher energies compared with the original response function; this behavior is not present in the Phys-NN reconstructions.  In the two-peak case, Phys-NN captures the EL peak in the low-energy region for both values of $\sigma$. On the other hand, the MaxEnt reconstruction, already not fully satisfactory for $\sigma=10^{-4}$, fails to reproduce the EL peak for $\sigma=10^{-3}$. As with the one-peak case, for this higher noise level MaxEnt($10^{-3}$) overestimates the height QE peak, and its position is shifted toward higher energies than in the original response function. 

\begin{figure}[t]
    \centering
    \includegraphics[width = \columnwidth]{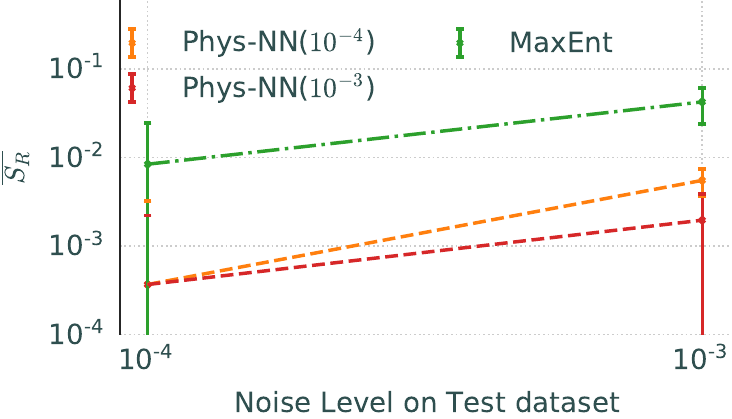}
    \caption{Change in the entropy from increasing the standard deviation of the Gaussian noise in the input Euclidean responses from $\sigma=10^{-4}$ to $\sigma=10^{-3}$.}
    \label{fig:change}
\end{figure}

To further quantify these results, we calculate the change in the entropy due to the increase in the noise level in the input in the test dataset. The average values of $S_R$ obtained from Phys-NN and MaxEnt calculations are plotted in Fig.~\ref{fig:change}. We observe that the change in the entropy due to the increase in the noise level is one order of magnitude larger for MaxEnt than that for  Phys-NN(${10^{-3}}$). 
%To further test the robustness of Phys-NN, we 
In Fig.~\ref{fig:change}, we also report results for Phys-NN(${10^{-4}}$), obtained by training Phys-NN on the low-noise data.
%and testing on the high-noise test dataset. 
In this case, the entropies increase by $4.00 \times10^{-4}$ and $51.3 \times10^{-4}$ for the one-peak and two-peak test datasets, respectively. While still a significant improvement compared with MaxEnt, the results for Phys-NN(${10^{-4}}$) are not as good as those obtained by Phys-NN(${10^{-3}}$). 
We conclude that Phys-NN is able to capture the main characteristics of the response functions even from noisier Euclidean responses.
%, and it has proven to be more robust than MaxEnt against higher noise levels in the input. 
We note that it is beneficial to be able to train on a set of responses having noise levels comparable to those of the target Euclidean responses. 

\section{Conclusions}
\label{sec:conc}
This work introduces Phys-NN, a physics-informed ANN approach to  approximately invert the Laplace transform and reliably reconstruct the electromagnetic response functions of atomic nuclei from their corresponding Euclidean responses. 
%The first layer of Phys-NN uses a Gaussian basis expansion to capture the main features of the input Euclidean responses. The second layer connects these Gaussian units and passes them to an exponential function, which provides the output of the network. 

We train, validate, and test Phys-NN, using 1 million response functions that exhibit the same features as those measured in electron scattering experiments. Half of the simulated responses are characterized by a single asymmetric broad peak in the quasielastic region; the other half possess an additional sharp Gaussian peak to model the low-energy transfer elastic transition. Unbiased Gaussian noise ($\sigma=10^{-4}$) is added to the Euclidean responses to simulate  the  statistical error of typical GFMC calculations for the $^4$He nucleus.
%with $\sigma=10^{-4}$ 
For training, we use a loss function with two terms. The first, inspired by the MaxEnt method, is an entropic loss to keep the reconstructed response functions close to the original ones. To avoid flat directions and improve the convergence of the optimization, we include a second term that seeks to keep the Laplace transform of the reconstructed responses close to the input Euclidean responses.

On a test dataset independent of that used in the training, 
%Using a test dataset comprised of $2,000$ pairs of response functions and Euclidean responses 
we demonstrate that Phys-NN significantly outperforms MaxEnt in terms of both the $S_R$ and $1-R_R^2$ metrics, especially on response functions characterized by two peaks. 
%Being based on the minimization of $\chi^2_E$, MaxEnt provides values of $\chi^2_E$ that are consistently closer to one than the Phys-NN. However, because of the ill-posed nature of the inversion problem, we demonstrate with scatter plots that achieving $\chi^2_E\approx 1$ does not necessarily imply good reconstruction performances. To make our analysis more explicit, we compare the best, average, and worst Phys-NN one-peak and two-peaks reconstructed response functions with MaxEnt predictions. 
Direct examination of the reconstructed responses shows that Phys-NN is capable of capturing the low-energy structures of the responses that are often completely missed by MaxEnt. We also find that Phys-NN better reproduces the position and height of the QE peak. 
%To further corroborate our analysis, we introduce 
Phys-NN produces about an order of magnitude improvement over MaxEnt in an energy-dependent entropy measure, especially for energy transfer up to $200$ MeV. This feature of Phys-NN is  promising for the reliable reconstruction of the low-energy structure of nuclear response functions and muon capture rates from GFMC calculations of the Euclidean responses. 
%We demonstrate that the Phys-NN is more robust than MaxEnt against increasing noise level in the input to $\sigma=10^{-3}$, a value compatible with typical GFMC calculations of $^{12}$C. We find that the change in the $S_R$ to due the higher noise level is one order of magnitude larger for the MaxEnt method than for the Phys-NN.

Our results show that Phys-NN is robust on a number of levels. First, Phys-NN has only two hyperparameters (the number of ANN Gaussians and the learning rate), and the relatively small amount of validation data used for determining values for these proved to be sufficient. Second, the Phys-NN outputs from ten independent training trials show remarkably little spread among the predicted responses, indicating a desirable insensitivity within the training process employed. 
%To gauge the sensitivity of the Phys-NN results to different initialization of the neural network, we performed ten independent training procedures, using distinct random seeds to get the initial parameters. In the reconstructed responses we observe a small spread among the ten realizations of the Phys-NN, corroborating the robustness of our method. 
We stress that the associated uncertainty bands do not represent the full theoretical error of our predictions, which in principle requires propagating the statistical errors of the Euclidean response through the response functions. In future work, we intend to include full uncertainty quantification and propagation by leveraging the linearity of the Laplace transform.
Third, when deployed on noisier testing data, Phys-NN 
%does not exhibit notable deterioration in  performance.  
maintains its advantage over MaxEnt.

In addition to the Laplace transform, primarily utilized within the GFMC method, the Lorentz kernel is  commonly used in the nuclear physics community~\cite{Efros:1994iq}. While initially restricted to light nuclear systems~\cite{Efros:1997jf,Bacca:2001kr,Bacca:2008tb}, its domain of applicability has recently been extended to study electron-nucleus interactions of medium-mass nuclei~\cite{Bacca:2013dma,Birkhan:2016qkr,Simonis:2019spj}. Similarly the Gaussian kernel has been found to be applicable in quantum algorithms with near-optimal computational cost to study the problem of spectral density estimation~\cite{Roggero:2020qoz}. We plan on generalizing the Phys-NN method to accommodate the inversion of both the Lorentz and Gaussian kernels, with the goal of improving  existing techniques. 

\acknowledgments
%Thanks for funding:
This work was supported in part by the U.S.\ Department of Energy (DOE), Office of Science, Offices of Advanced Scientific Computing Research and Nuclear Physics, by the Argonne LDRD program, and by the NUCLEI, FASTMath, and RAPIDS SciDAC projects under contract number DE-AC02-06CH11357. 
N.R.\ was also supported by Fermi Research Alliance, LLC under contract number DE-AC02-07CH11359 with the U.S.\ DOE, Office of Science, Office of High Energy Physics.
S.M.W.\ was also supported by the National Science Foundation CSSI program under award number OAC-2004601 (BAND Collaboration). 
P.B., A.L., and S.M.W.\ were also supported by DOE Early Career Research Program awards.
%SW: do we need to acknowledge any ALCF resources?
%AL: I think that Krishnan used primarily JLSE, but I might be wrong
We are grateful for the computing resources from the Joint Laboratory for System Evaluation  and Leadership Computing Facility at Argonne.
%Dropped the NL since it was dropped earlier in acknowledgments.

% \appendix
% \section{Metrics and their definition.}

% \subsection{$\tilde{\chi}^{2}$ -- a modified $\chi^2$} 
% From Eq. \eqref{eq:chi2}, the 
% $E(\tau_i) = \sum_j K(\tau_i, \omega_j) R(\omega_j)$
% where $K(\tau_i, \omega_j) = \exp(-\tau_i \omega_j) * d\omega$
% By doing the naive error propagation, we get
% $\sigma_E(\tau_i) = \sqrt{ \sum_j K(\tau_i, \omega_j)^2 \sigma_R(\omega_j)^2 }$
% this implies that the ``average'' Euclidean error is given by
% $$< \sigma_E > = 1 / n_\tau \sum_i \sqrt{ \sum_j K(\tau_i, \omega_j)^2 \sigma_R (\omega_j)^2 }$$
% $$< \sigma_E > = 1 / n_\tau \sum_i \sqrt{ ( \sum_j K(\tau_i, \omega_j)^2 ) <\sigma_R >}$$
% where in the last line I defined the ``average" response error $<\sigma_R >$ . The ``conversion'' factor that we need to compute is
% $$\frac{1}{n_{\tau}} \sum_i \sqrt{ \sum_j K(\tau_i, \omega_j)^2}$$

% \subsection{$R^{2}$}
%%%%%%%%%%%%%%%%%%%%%%%%%%%%%%%%%%%%%%%%%%%%%%%%%%%%%%%%%%%%%%%%%%%%%%%%%%%%%%%%%%%%%
%%%%%%%%%%%%%%%%%%%%%%%%%%%%%%%%%%%%%%%%%%%%%%%%%%%%%%%%%%%%%%%%%%%%%%%%%%%%%%%%%%%%%
%%%%%%%%%%%%%%%%%%%%%%%%%%%%%%%%%%%%%%%%%%%%%%%%%%%%%%%%%%%%%%%%%%%%%%%%%%%%%%%%%%%%%
%%%%%%%%%%%%%%%%%%%%%%%%%%%%%%%%%%%%%%%%%%%%%%%%%%%%%%%%%%%%%%%%%%%%%%%%%%%%%%%%%%%%%

\section*{Bibliography}

\bibliography{bib/biblio}

\begin{thebibliography}{55}%
\makeatletter
\providecommand \@ifxundefined [1]{%
 \@ifx{#1\undefined}
}%
\providecommand \@ifnum [1]{%
 \ifnum #1\expandafter \@firstoftwo
 \else \expandafter \@secondoftwo
 \fi
}%
\providecommand \@ifx [1]{%
 \ifx #1\expandafter \@firstoftwo
 \else \expandafter \@secondoftwo
 \fi
}%
\providecommand \natexlab [1]{#1}%
\providecommand \enquote  [1]{``#1''}%
\providecommand \bibnamefont  [1]{#1}%
\providecommand \bibfnamefont [1]{#1}%
\providecommand \citenamefont [1]{#1}%
\providecommand \href@noop [0]{\@secondoftwo}%
\providecommand \href [0]{\begingroup \@sanitize@url \@href}%
\providecommand \@href[1]{\@@startlink{#1}\@@href}%
\providecommand \@@href[1]{\endgroup#1\@@endlink}%
\providecommand \@sanitize@url [0]{\catcode `\\12\catcode `\$12\catcode
  `\&12\catcode `\#12\catcode `\^12\catcode `\_12\catcode `\%12\relax}%
\providecommand \@@startlink[1]{}%
\providecommand \@@endlink[0]{}%
\providecommand \url  [0]{\begingroup\@sanitize@url \@url }%
\providecommand \@url [1]{\endgroup\@href {#1}{\urlprefix }}%
\providecommand \urlprefix  [0]{URL }%
\providecommand \Eprint [0]{\href }%
\providecommand \doibase [0]{http://dx.doi.org/}%
\providecommand \selectlanguage [0]{\@gobble}%
\providecommand \bibinfo  [0]{\@secondoftwo}%
\providecommand \bibfield  [0]{\@secondoftwo}%
\providecommand \translation [1]{[#1]}%
\providecommand \BibitemOpen [0]{}%
\providecommand \bibitemStop [0]{}%
\providecommand \bibitemNoStop [0]{.\EOS\space}%
\providecommand \EOS [0]{\spacefactor3000\relax}%
\providecommand \BibitemShut  [1]{\csname bibitem#1\endcsname}%
\let\auto@bib@innerbib\@empty
%</preamble>
\bibitem [{\citenamefont {Benhar}\ \emph {et~al.}(2008)\citenamefont {Benhar},
  \citenamefont {Day},\ and\ \citenamefont {Sick}}]{Benhar:2006wy}%
  \BibitemOpen
  \bibfield  {author} {\bibinfo {author} {\bibfnamefont {O.}~\bibnamefont
  {Benhar}}, \bibinfo {author} {\bibfnamefont {D.}~\bibnamefont {Day}}, \ and\
  \bibinfo {author} {\bibfnamefont {I.}~\bibnamefont {Sick}},\ }\href {\doibase
  10.1103/RevModPhys.80.189} {\bibfield  {journal} {\bibinfo  {journal} {Rev.\
  Mod.\ Phys.}\ }\textbf {\bibinfo {volume} {80}},\ \bibinfo {pages} {189}
  (\bibinfo {year} {2008})},\ \Eprint {http://arxiv.org/abs/nucl-ex/0603029}
  {arXiv:nucl-ex/0603029} \BibitemShut {NoStop}%
\bibitem [{\citenamefont {Wiringa}\ \emph {et~al.}(2014)\citenamefont
  {Wiringa}, \citenamefont {Schiavilla}, \citenamefont {Pieper},\ and\
  \citenamefont {Carlson}}]{Wiringa:2013ala}%
  \BibitemOpen
  \bibfield  {author} {\bibinfo {author} {\bibfnamefont {R.}~\bibnamefont
  {Wiringa}}, \bibinfo {author} {\bibfnamefont {R.}~\bibnamefont {Schiavilla}},
  \bibinfo {author} {\bibfnamefont {S.~C.}\ \bibnamefont {Pieper}}, \ and\
  \bibinfo {author} {\bibfnamefont {J.}~\bibnamefont {Carlson}},\ }\href
  {\doibase 10.1103/PhysRevC.89.024305} {\bibfield  {journal} {\bibinfo
  {journal} {Phys. Rev. C}\ }\textbf {\bibinfo {volume} {89}},\ \bibinfo
  {pages} {024305} (\bibinfo {year} {2014})},\ \Eprint
  {http://arxiv.org/abs/1309.3794} {arXiv:1309.3794 [nucl-th]} \BibitemShut
  {NoStop}%
\bibitem [{\citenamefont {Ciofi~degli Atti}(2015)}]{Atti:2015eda}%
  \BibitemOpen
  \bibfield  {author} {\bibinfo {author} {\bibfnamefont {C.}~\bibnamefont
  {Ciofi~degli Atti}},\ }\href {\doibase 10.1016/j.physrep.2015.06.002}
  {\bibfield  {journal} {\bibinfo  {journal} {Phys. Rept.}\ }\textbf {\bibinfo
  {volume} {590}},\ \bibinfo {pages} {1} (\bibinfo {year} {2015})}\BibitemShut
  {NoStop}%
%%CITATION = PRPLC,590,1;%%
\bibitem [{\citenamefont {Hen}\ \emph {et~al.}(2017)\citenamefont {Hen},
  \citenamefont {Miller}, \citenamefont {Piasetzky},\ and\ \citenamefont
  {Weinstein}}]{Hen:2016kwk}%
  \BibitemOpen
  \bibfield  {author} {\bibinfo {author} {\bibfnamefont {O.}~\bibnamefont
  {Hen}}, \bibinfo {author} {\bibfnamefont {G.~A.}\ \bibnamefont {Miller}},
  \bibinfo {author} {\bibfnamefont {E.}~\bibnamefont {Piasetzky}}, \ and\
  \bibinfo {author} {\bibfnamefont {L.~B.}\ \bibnamefont {Weinstein}},\ }\href
  {\doibase 10.1103/RevModPhys.89.045002} {\bibfield  {journal} {\bibinfo
  {journal} {Rev. Mod. Phys.}\ }\textbf {\bibinfo {volume} {89}},\ \bibinfo
  {pages} {045002} (\bibinfo {year} {2017})},\ \Eprint
  {http://arxiv.org/abs/1611.09748} {arXiv:1611.09748 [nucl-ex]} \BibitemShut
  {NoStop}%
%%CITATION = ARXIV:1611.09748;%%
\bibitem [{\citenamefont {Amaro}\ \emph {et~al.}(2019)\citenamefont {Amaro},
  \citenamefont {Barbaro}, \citenamefont {Caballero}, \citenamefont
  {Gonz\'alez-Jim\'enez}, \citenamefont {Megias},\ and\ \citenamefont
  {Ruiz~Simo}}]{Amaro:2019zos}%
  \BibitemOpen
  \bibfield  {author} {\bibinfo {author} {\bibfnamefont {J.~E.}\ \bibnamefont
  {Amaro}}, \bibinfo {author} {\bibfnamefont {M.~B.}\ \bibnamefont {Barbaro}},
  \bibinfo {author} {\bibfnamefont {J.~A.}\ \bibnamefont {Caballero}}, \bibinfo
  {author} {\bibfnamefont {R.}~\bibnamefont {Gonz\'alez-Jim\'enez}}, \bibinfo
  {author} {\bibfnamefont {G.~D.}\ \bibnamefont {Megias}}, \ and\ \bibinfo
  {author} {\bibfnamefont {I.}~\bibnamefont {Ruiz~Simo}},\ }\href@noop {} {\
  (\bibinfo {year} {2019})},\ \Eprint {http://arxiv.org/abs/1912.10612}
  {arXiv:1912.10612 [nucl-th]} \BibitemShut {NoStop}%
%%CITATION = ARXIV:1912.10612;%%
\bibitem [{The {NO}v{A} {E}xperiment()}]{nova_web}%
  \BibitemOpen
  The {NO}v{A} {E}xperiment,\ \href@noop {} {}\bibinfo {howpublished}
  {\url{http://www-nova.fnal.gov}}\BibitemShut {NoStop}%
\bibitem [{The {T}2{K} {E}xperiment()}]{t2k_web}%
  \BibitemOpen
  The {T}2{K} {E}xperiment,\ \href@noop {} {}\bibinfo {howpublished}
  {\url{http://t2k-experiment.org}}\BibitemShut {NoStop}%
\bibitem [{The Deep Underground Neutrino Experiment()}]{dune_web}%
  \BibitemOpen
  The Deep Underground Neutrino Experiment,\ \href@noop {} {}\bibinfo
  {howpublished} {\url{http://www.dunescience.org}}\BibitemShut {NoStop}%
\bibitem [{Hyper-Kamiokande()}]{hk_web}%
  \BibitemOpen
  Hyper-Kamiokande,\ \href@noop {} {}\bibinfo {howpublished}
  {\url{http://www.hyperk.org}}\BibitemShut {NoStop}%
\bibitem [{\citenamefont {Benhar}\ \emph {et~al.}(2017)\citenamefont {Benhar},
  \citenamefont {Huber}, \citenamefont {Mariani},\ and\ \citenamefont
  {Meloni}}]{Benhar:2015wva}%
  \BibitemOpen
  \bibfield  {author} {\bibinfo {author} {\bibfnamefont {O.}~\bibnamefont
  {Benhar}}, \bibinfo {author} {\bibfnamefont {P.}~\bibnamefont {Huber}},
  \bibinfo {author} {\bibfnamefont {C.}~\bibnamefont {Mariani}}, \ and\
  \bibinfo {author} {\bibfnamefont {D.}~\bibnamefont {Meloni}},\ }\href
  {\doibase 10.1016/j.physrep.2017.07.004} {\bibfield  {journal} {\bibinfo
  {journal} {Phys. Rept.}\ }\textbf {\bibinfo {volume} {700}},\ \bibinfo
  {pages} {1} (\bibinfo {year} {2017})},\ \Eprint
  {http://arxiv.org/abs/1501.06448} {arXiv:1501.06448 [nucl-th]} \BibitemShut
  {NoStop}%
%%CITATION = ARXIV:1501.06448;%%
\bibitem [{\citenamefont {Katori}\ and\ \citenamefont
  {Martini}(2018)}]{Katori:2016yel}%
  \BibitemOpen
  \bibfield  {author} {\bibinfo {author} {\bibfnamefont {T.}~\bibnamefont
  {Katori}}\ and\ \bibinfo {author} {\bibfnamefont {M.}~\bibnamefont
  {Martini}},\ }\href {\doibase 10.1088/1361-6471/aa8bf7} {\bibfield  {journal}
  {\bibinfo  {journal} {J. Phys. G}\ }\textbf {\bibinfo {volume} {45}},\
  \bibinfo {pages} {013001} (\bibinfo {year} {2018})},\ \Eprint
  {http://arxiv.org/abs/1611.07770} {arXiv:1611.07770 [hep-ph]} \BibitemShut
  {NoStop}%
%%CITATION = ARXIV:1611.07770;%%
\bibitem [{\citenamefont {Alvarez-Ruso}\ \emph {et~al.}(2018)\citenamefont
  {Alvarez-Ruso} \emph {et~al.}}]{Alvarez-Ruso:2017oui}%
  \BibitemOpen
  \bibfield  {author} {\bibinfo {author} {\bibfnamefont {L.}~\bibnamefont
  {Alvarez-Ruso}} \emph {et~al.} (\bibinfo {collaboration} {NuSTEC}),\ }\href
  {\doibase 10.1016/j.ppnp.2018.01.006} {\bibfield  {journal} {\bibinfo
  {journal} {Prog. Part. Nucl. Phys.}\ }\textbf {\bibinfo {volume} {100}},\
  \bibinfo {pages} {1} (\bibinfo {year} {2018})},\ \Eprint
  {http://arxiv.org/abs/1706.03621} {arXiv:1706.03621 [hep-ph]} \BibitemShut
  {NoStop}%
%%CITATION = ARXIV:1706.03621;%%
\bibitem [{\citenamefont {Acciarri}\ \emph {et~al.}(2014)\citenamefont
  {Acciarri} \emph {et~al.}}]{Acciarri:2014gev}%
  \BibitemOpen
  \bibfield  {author} {\bibinfo {author} {\bibfnamefont {R.}~\bibnamefont
  {Acciarri}} \emph {et~al.} (\bibinfo {collaboration} {ArgoNeuT}),\ }\href
  {\doibase 10.1103/PhysRevD.90.012008} {\bibfield  {journal} {\bibinfo
  {journal} {Phys. Rev. D}\ }\textbf {\bibinfo {volume} {90}},\ \bibinfo
  {pages} {012008} (\bibinfo {year} {2014})},\ \Eprint
  {http://arxiv.org/abs/1405.4261} {arXiv:1405.4261 [nucl-ex]} \BibitemShut
  {NoStop}%
\bibitem [{\citenamefont {Carlson}\ \emph {et~al.}(2015)\citenamefont
  {Carlson}, \citenamefont {Gandolfi}, \citenamefont {Pederiva}, \citenamefont
  {Pieper}, \citenamefont {Schiavilla}, \citenamefont {Schmidt},\ and\
  \citenamefont {Wiringa}}]{Carlson:2014vla}%
  \BibitemOpen
  \bibfield  {author} {\bibinfo {author} {\bibfnamefont {J.}~\bibnamefont
  {Carlson}}, \bibinfo {author} {\bibfnamefont {S.}~\bibnamefont {Gandolfi}},
  \bibinfo {author} {\bibfnamefont {F.}~\bibnamefont {Pederiva}}, \bibinfo
  {author} {\bibfnamefont {S.~C.}\ \bibnamefont {Pieper}}, \bibinfo {author}
  {\bibfnamefont {R.}~\bibnamefont {Schiavilla}}, \bibinfo {author}
  {\bibfnamefont {K.}~\bibnamefont {Schmidt}}, \ and\ \bibinfo {author}
  {\bibfnamefont {R.}~\bibnamefont {Wiringa}},\ }\href {\doibase
  10.1103/RevModPhys.87.1067} {\bibfield  {journal} {\bibinfo  {journal} {Rev.
  Mod. Phys.}\ }\textbf {\bibinfo {volume} {87}},\ \bibinfo {pages} {1067}
  (\bibinfo {year} {2015})},\ \Eprint {http://arxiv.org/abs/1412.3081}
  {arXiv:1412.3081 [nucl-th]} \BibitemShut {NoStop}%
\bibitem [{\citenamefont {Carlson}\ \emph {et~al.}(2002)\citenamefont
  {Carlson}, \citenamefont {Jourdan}, \citenamefont {Schiavilla},\ and\
  \citenamefont {Sick}}]{Carlson:2001mp}%
  \BibitemOpen
  \bibfield  {author} {\bibinfo {author} {\bibfnamefont {J.}~\bibnamefont
  {Carlson}}, \bibinfo {author} {\bibfnamefont {J.}~\bibnamefont {Jourdan}},
  \bibinfo {author} {\bibfnamefont {R.}~\bibnamefont {Schiavilla}}, \ and\
  \bibinfo {author} {\bibfnamefont {I.}~\bibnamefont {Sick}},\ }\href {\doibase
  10.1103/PhysRevC.65.024002} {\bibfield  {journal} {\bibinfo  {journal} {Phys.
  Rev. C}\ }\textbf {\bibinfo {volume} {65}},\ \bibinfo {pages} {024002}
  (\bibinfo {year} {2002})},\ \Eprint {http://arxiv.org/abs/nucl-th/0106047}
  {arXiv:nucl-th/0106047} \BibitemShut {NoStop}%
\bibitem [{\citenamefont {Lovato}\ \emph {et~al.}(2016)\citenamefont {Lovato},
  \citenamefont {Gandolfi}, \citenamefont {Carlson}, \citenamefont {Pieper},\
  and\ \citenamefont {Schiavilla}}]{Lovato:2016gkq}%
  \BibitemOpen
  \bibfield  {author} {\bibinfo {author} {\bibfnamefont {A.}~\bibnamefont
  {Lovato}}, \bibinfo {author} {\bibfnamefont {S.}~\bibnamefont {Gandolfi}},
  \bibinfo {author} {\bibfnamefont {J.}~\bibnamefont {Carlson}}, \bibinfo
  {author} {\bibfnamefont {S.~C.}\ \bibnamefont {Pieper}}, \ and\ \bibinfo
  {author} {\bibfnamefont {R.}~\bibnamefont {Schiavilla}},\ }\href {\doibase
  10.1103/PhysRevLett.117.082501} {\bibfield  {journal} {\bibinfo  {journal}
  {Phys. Rev. Lett.}\ }\textbf {\bibinfo {volume} {117}},\ \bibinfo {pages}
  {082501} (\bibinfo {year} {2016})},\ \Eprint
  {http://arxiv.org/abs/1605.00248} {arXiv:1605.00248 [nucl-th]} \BibitemShut
  {NoStop}%
\bibitem [{\citenamefont {Lovato}\ \emph {et~al.}(2018)\citenamefont {Lovato},
  \citenamefont {Gandolfi}, \citenamefont {Carlson}, \citenamefont {Lusk},
  \citenamefont {Pieper},\ and\ \citenamefont {Schiavilla}}]{Lovato:2017cux}%
  \BibitemOpen
  \bibfield  {author} {\bibinfo {author} {\bibfnamefont {A.}~\bibnamefont
  {Lovato}}, \bibinfo {author} {\bibfnamefont {S.}~\bibnamefont {Gandolfi}},
  \bibinfo {author} {\bibfnamefont {J.}~\bibnamefont {Carlson}}, \bibinfo
  {author} {\bibfnamefont {E.}~\bibnamefont {Lusk}}, \bibinfo {author}
  {\bibfnamefont {S.~C.}\ \bibnamefont {Pieper}}, \ and\ \bibinfo {author}
  {\bibfnamefont {R.}~\bibnamefont {Schiavilla}},\ }\href {\doibase
  10.1103/PhysRevC.97.022502} {\bibfield  {journal} {\bibinfo  {journal} {Phys.
  Rev. C}\ }\textbf {\bibinfo {volume} {97}},\ \bibinfo {pages} {022502}
  (\bibinfo {year} {2018})},\ \Eprint {http://arxiv.org/abs/1711.02047}
  {arXiv:1711.02047 [nucl-th]} \BibitemShut {NoStop}%
\bibitem [{\citenamefont {Lovato}\ \emph {et~al.}(2020)\citenamefont {Lovato},
  \citenamefont {Carlson}, \citenamefont {Gandolfi}, \citenamefont {Rocco},\
  and\ \citenamefont {Schiavilla}}]{Lovato:2020kba}%
  \BibitemOpen
  \bibfield  {author} {\bibinfo {author} {\bibfnamefont {A.}~\bibnamefont
  {Lovato}}, \bibinfo {author} {\bibfnamefont {J.}~\bibnamefont {Carlson}},
  \bibinfo {author} {\bibfnamefont {S.}~\bibnamefont {Gandolfi}}, \bibinfo
  {author} {\bibfnamefont {N.}~\bibnamefont {Rocco}}, \ and\ \bibinfo {author}
  {\bibfnamefont {R.}~\bibnamefont {Schiavilla}},\ }\href@noop {} {\  (\bibinfo
  {year} {2020})},\ \Eprint {http://arxiv.org/abs/2003.07710} {arXiv:2003.07710
  [nucl-th]} \BibitemShut {NoStop}%
\bibitem [{\citenamefont {Lovato}\ \emph {et~al.}(2019)\citenamefont {Lovato},
  \citenamefont {Rocco},\ and\ \citenamefont {Schiavilla}}]{Lovato:2019fiw}%
  \BibitemOpen
  \bibfield  {author} {\bibinfo {author} {\bibfnamefont {A.}~\bibnamefont
  {Lovato}}, \bibinfo {author} {\bibfnamefont {N.}~\bibnamefont {Rocco}}, \
  and\ \bibinfo {author} {\bibfnamefont {R.}~\bibnamefont {Schiavilla}},\
  }\href {\doibase 10.1103/PhysRevC.100.035502} {\bibfield  {journal} {\bibinfo
   {journal} {Phys. Rev. C}\ }\textbf {\bibinfo {volume} {100}},\ \bibinfo
  {pages} {035502} (\bibinfo {year} {2019})},\ \Eprint
  {http://arxiv.org/abs/1903.08078} {arXiv:1903.08078 [nucl-th]} \BibitemShut
  {NoStop}%
\bibitem [{\citenamefont {Bryan}(1990)}]{Bryan:1990}%
  \BibitemOpen
  \bibfield  {author} {\bibinfo {author} {\bibfnamefont {R.}~\bibnamefont
  {Bryan}},\ }\href {\doibase 10.1007/BF02427376} {\bibfield  {journal}
  {\bibinfo  {journal} {Eur. Biophys. J.}\ }\textbf {\bibinfo {volume} {18}},\
  \bibinfo {pages} {165} (\bibinfo {year} {1990})}\BibitemShut {NoStop}%
\bibitem [{\citenamefont {Jarrell}\ and\ \citenamefont
  {Gubernatis}(1996)}]{Jarrell:1996}%
  \BibitemOpen
  \bibfield  {author} {\bibinfo {author} {\bibfnamefont {M.}~\bibnamefont
  {Jarrell}}\ and\ \bibinfo {author} {\bibfnamefont {J.}~\bibnamefont
  {Gubernatis}},\ }\href {\doibase 10.1016/0370-1573(95)00074-7} {\bibfield
  {journal} {\bibinfo  {journal} {Phys. Rept.}\ }\textbf {\bibinfo {volume}
  {269}},\ \bibinfo {pages} {133} (\bibinfo {year} {1996})}\BibitemShut
  {NoStop}%
\bibitem [{\citenamefont {Carleo}\ \emph {et~al.}(2019)\citenamefont {Carleo},
  \citenamefont {Cirac}, \citenamefont {Cranmer}, \citenamefont {Daudet},
  \citenamefont {Schuld}, \citenamefont {Tishby}, \citenamefont
  {Vogt-Maranto},\ and\ \citenamefont {Zdeborov\'a}}]{Carleo:2019}%
  \BibitemOpen
  \bibfield  {author} {\bibinfo {author} {\bibfnamefont {G.}~\bibnamefont
  {Carleo}}, \bibinfo {author} {\bibfnamefont {I.}~\bibnamefont {Cirac}},
  \bibinfo {author} {\bibfnamefont {K.}~\bibnamefont {Cranmer}}, \bibinfo
  {author} {\bibfnamefont {L.}~\bibnamefont {Daudet}}, \bibinfo {author}
  {\bibfnamefont {M.}~\bibnamefont {Schuld}}, \bibinfo {author} {\bibfnamefont
  {N.}~\bibnamefont {Tishby}}, \bibinfo {author} {\bibfnamefont
  {L.}~\bibnamefont {Vogt-Maranto}}, \ and\ \bibinfo {author} {\bibfnamefont
  {L.}~\bibnamefont {Zdeborov\'a}},\ }\href {\doibase
  10.1103/RevModPhys.91.045002} {\bibfield  {journal} {\bibinfo  {journal}
  {Rev. Mod. Phys.}\ }\textbf {\bibinfo {volume} {91}},\ \bibinfo {pages}
  {045002} (\bibinfo {year} {2019})}\BibitemShut {NoStop}%
\bibitem [{\citenamefont {Negoita}\ \emph {et~al.}(2019)\citenamefont
  {Negoita}, \citenamefont {Vary}, \citenamefont {Luecke}, \citenamefont
  {Maris}, \citenamefont {Shirokov}, \citenamefont {Shin}, \citenamefont {Kim},
  \citenamefont {Ng}, \citenamefont {Yang}, \citenamefont {Lockner},\ and\
  \citenamefont {Prabhu}}]{Negoita2019}%
  \BibitemOpen
  \bibfield  {author} {\bibinfo {author} {\bibfnamefont {G.~A.}\ \bibnamefont
  {Negoita}}, \bibinfo {author} {\bibfnamefont {J.~P.}\ \bibnamefont {Vary}},
  \bibinfo {author} {\bibfnamefont {G.~R.}\ \bibnamefont {Luecke}}, \bibinfo
  {author} {\bibfnamefont {P.}~\bibnamefont {Maris}}, \bibinfo {author}
  {\bibfnamefont {A.~M.}\ \bibnamefont {Shirokov}}, \bibinfo {author}
  {\bibfnamefont {I.~J.}\ \bibnamefont {Shin}}, \bibinfo {author}
  {\bibfnamefont {Y.}~\bibnamefont {Kim}}, \bibinfo {author} {\bibfnamefont
  {E.~G.}\ \bibnamefont {Ng}}, \bibinfo {author} {\bibfnamefont
  {C.}~\bibnamefont {Yang}}, \bibinfo {author} {\bibfnamefont {M.}~\bibnamefont
  {Lockner}}, \ and\ \bibinfo {author} {\bibfnamefont {G.~M.}\ \bibnamefont
  {Prabhu}},\ }\href {\doibase 10.1103/physrevc.99.054308} {\bibfield
  {journal} {\bibinfo  {journal} {Phys. Rev. C}\ }\textbf {\bibinfo {volume}
  {99}} (\bibinfo {year} {2019}),\ 10.1103/physrevc.99.054308}\BibitemShut
  {NoStop}%
\bibitem [{\citenamefont {Jiang}\ \emph {et~al.}(2019)\citenamefont {Jiang},
  \citenamefont {Hagen},\ and\ \citenamefont {Papenbrock}}]{Jiang2019}%
  \BibitemOpen
  \bibfield  {author} {\bibinfo {author} {\bibfnamefont {W.~G.}\ \bibnamefont
  {Jiang}}, \bibinfo {author} {\bibfnamefont {G.}~\bibnamefont {Hagen}}, \ and\
  \bibinfo {author} {\bibfnamefont {T.}~\bibnamefont {Papenbrock}},\ }\href
  {\doibase 10.1103/physrevc.100.054326} {\bibfield  {journal} {\bibinfo
  {journal} {Phys. Rev. C}\ }\textbf {\bibinfo {volume} {100}} (\bibinfo {year}
  {2019}),\ 10.1103/physrevc.100.054326}\BibitemShut {NoStop}%
\bibitem [{\citenamefont {Neufcourt}\ \emph {et~al.}(2019)\citenamefont
  {Neufcourt}, \citenamefont {Cao}, \citenamefont {Nazarewicz}, \citenamefont
  {Olsen},\ and\ \citenamefont {Viens}}]{Neufcourt2019}%
  \BibitemOpen
  \bibfield  {author} {\bibinfo {author} {\bibfnamefont {L.}~\bibnamefont
  {Neufcourt}}, \bibinfo {author} {\bibfnamefont {Y.}~\bibnamefont {Cao}},
  \bibinfo {author} {\bibfnamefont {W.}~\bibnamefont {Nazarewicz}}, \bibinfo
  {author} {\bibfnamefont {E.}~\bibnamefont {Olsen}}, \ and\ \bibinfo {author}
  {\bibfnamefont {F.}~\bibnamefont {Viens}},\ }\href {\doibase
  10.1103/PhysRevLett.122.062502} {\bibfield  {journal} {\bibinfo  {journal}
  {Phys. Rev. Lett.}\ }\textbf {\bibinfo {volume} {122}},\ \bibinfo {pages}
  {062502} (\bibinfo {year} {2019})}\BibitemShut {NoStop}%
\bibitem [{\citenamefont {Keeble}\ and\ \citenamefont
  {Rios}(2020)}]{Keeble:2019bkv}%
  \BibitemOpen
  \bibfield  {author} {\bibinfo {author} {\bibfnamefont {J.}~\bibnamefont
  {Keeble}}\ and\ \bibinfo {author} {\bibfnamefont {A.}~\bibnamefont {Rios}},\
  }\href {\doibase 10.1016/j.physletb.2020.135743} {\bibfield  {journal}
  {\bibinfo  {journal} {Phys. Lett. B}\ }\textbf {\bibinfo {volume} {809}},\
  \bibinfo {pages} {135743} (\bibinfo {year} {2020})},\ \Eprint
  {http://arxiv.org/abs/1911.13092} {arXiv:1911.13092 [nucl-th]} \BibitemShut
  {NoStop}%
\bibitem [{\citenamefont {Adams}\ \emph {et~al.}(2020)\citenamefont {Adams},
  \citenamefont {Carleo}, \citenamefont {Lovato},\ and\ \citenamefont
  {Rocco}}]{Adams:2020aax}%
  \BibitemOpen
  \bibfield  {author} {\bibinfo {author} {\bibfnamefont {C.}~\bibnamefont
  {Adams}}, \bibinfo {author} {\bibfnamefont {G.}~\bibnamefont {Carleo}},
  \bibinfo {author} {\bibfnamefont {A.}~\bibnamefont {Lovato}}, \ and\ \bibinfo
  {author} {\bibfnamefont {N.}~\bibnamefont {Rocco}},\ }\href@noop {} {\
  (\bibinfo {year} {2020})},\ \Eprint {http://arxiv.org/abs/2007.14282}
  {arXiv:2007.14282 [nucl-th]} \BibitemShut {NoStop}%
\bibitem [{\citenamefont {{McCann}}\ \emph {et~al.}(2017)\citenamefont
  {{McCann}}, \citenamefont {{Jin}},\ and\ \citenamefont
  {{Unser}}}]{McCann:2017}%
  \BibitemOpen
  \bibfield  {author} {\bibinfo {author} {\bibfnamefont {M.~T.}\ \bibnamefont
  {{McCann}}}, \bibinfo {author} {\bibfnamefont {K.~H.}\ \bibnamefont {{Jin}}},
  \ and\ \bibinfo {author} {\bibfnamefont {M.}~\bibnamefont {{Unser}}},\ }\href
  {\doibase 10.1109/MSP.2017.2739299} {\bibfield  {journal} {\bibinfo
  {journal} {IEEE Signal Process. Mag.}\ }\textbf {\bibinfo {volume} {34}},\
  \bibinfo {pages} {85} (\bibinfo {year} {2017})},\ \Eprint
  {http://arxiv.org/abs/1710.04011} {arXiv:1710.04011 [eess.IV]} \BibitemShut
  {NoStop}%
\bibitem [{\citenamefont {{Arsenault}}\ \emph {et~al.}(2017)\citenamefont
  {{Arsenault}}, \citenamefont {{Neuberg}}, \citenamefont {{Hannah}},\ and\
  \citenamefont {{Millis}}}]{Arsenault:2017}%
  \BibitemOpen
  \bibfield  {author} {\bibinfo {author} {\bibfnamefont {L.-F.}\ \bibnamefont
  {{Arsenault}}}, \bibinfo {author} {\bibfnamefont {R.}~\bibnamefont
  {{Neuberg}}}, \bibinfo {author} {\bibfnamefont {L.~A.}\ \bibnamefont
  {{Hannah}}}, \ and\ \bibinfo {author} {\bibfnamefont {A.~J.}\ \bibnamefont
  {{Millis}}},\ }\href {\doibase 10.1088/1361-6420/aa8d93} {\bibfield
  {journal} {\bibinfo  {journal} {Inverse Probl.}\ }\textbf {\bibinfo {volume}
  {33}},\ \bibinfo {eid} {115007} (\bibinfo {year} {2017})}\BibitemShut
  {NoStop}%
\bibitem [{\citenamefont {{Yoon}}\ \emph {et~al.}(2018)\citenamefont {{Yoon}},
  \citenamefont {{Sim}},\ and\ \citenamefont {{Han}}}]{Yoon:2018}%
  \BibitemOpen
  \bibfield  {author} {\bibinfo {author} {\bibfnamefont {H.}~\bibnamefont
  {{Yoon}}}, \bibinfo {author} {\bibfnamefont {J.-H.}\ \bibnamefont {{Sim}}}, \
  and\ \bibinfo {author} {\bibfnamefont {M.~J.}\ \bibnamefont {{Han}}},\ }\href
  {\doibase 10.1103/PhysRevB.98.245101} {\bibfield  {journal} {\bibinfo
  {journal} {\prb}\ }\textbf {\bibinfo {volume} {98}},\ \bibinfo {eid} {245101}
  (\bibinfo {year} {2018})},\ \Eprint {http://arxiv.org/abs/1806.03841}
  {arXiv:1806.03841 [cond-mat.str-el]} \BibitemShut {NoStop}%
\bibitem [{\citenamefont {{Fournier}}\ \emph {et~al.}(2020)\citenamefont
  {{Fournier}}, \citenamefont {{Wang}}, \citenamefont {{Yazyev}},\ and\
  \citenamefont {{Wu}}}]{Fournier:2020}%
  \BibitemOpen
  \bibfield  {author} {\bibinfo {author} {\bibfnamefont {R.}~\bibnamefont
  {{Fournier}}}, \bibinfo {author} {\bibfnamefont {L.}~\bibnamefont {{Wang}}},
  \bibinfo {author} {\bibfnamefont {O.~V.}\ \bibnamefont {{Yazyev}}}, \ and\
  \bibinfo {author} {\bibfnamefont {Q.}~\bibnamefont {{Wu}}},\ }\href {\doibase
  10.1103/PhysRevLett.124.056401} {\bibfield  {journal} {\bibinfo  {journal}
  {\prl}\ }\textbf {\bibinfo {volume} {124}},\ \bibinfo {eid} {056401}
  (\bibinfo {year} {2020})}\BibitemShut {NoStop}%
\bibitem [{\citenamefont {{Xie}}\ \emph {et~al.}(2019)\citenamefont {{Xie}},
  \citenamefont {{Bao}}, \citenamefont {{Maier}},\ and\ \citenamefont
  {{Webster}}}]{Xie:2019}%
  \BibitemOpen
  \bibfield  {author} {\bibinfo {author} {\bibfnamefont {X.}~\bibnamefont
  {{Xie}}}, \bibinfo {author} {\bibfnamefont {F.}~\bibnamefont {{Bao}}},
  \bibinfo {author} {\bibfnamefont {T.}~\bibnamefont {{Maier}}}, \ and\
  \bibinfo {author} {\bibfnamefont {C.}~\bibnamefont {{Webster}}},\ }\href@noop
  {} {\bibfield  {journal} {\bibinfo  {journal} {arXiv}\ } (\bibinfo {year}
  {2019})},\ \Eprint {http://arxiv.org/abs/1905.10430} {arXiv:1905.10430
  [physics.comp-ph]} \BibitemShut {NoStop}%
\bibitem [{\citenamefont {Schmidt}\ and\ \citenamefont
  {Fantoni}(1999)}]{Schmidt:1999lik}%
  \BibitemOpen
  \bibfield  {author} {\bibinfo {author} {\bibfnamefont {K.}~\bibnamefont
  {Schmidt}}\ and\ \bibinfo {author} {\bibfnamefont {S.}~\bibnamefont
  {Fantoni}},\ }\href {\doibase 10.1016/S0370-2693(98)01522-6} {\bibfield
  {journal} {\bibinfo  {journal} {Phys. Lett. B}\ }\textbf {\bibinfo {volume}
  {446}},\ \bibinfo {pages} {99} (\bibinfo {year} {1999})}\BibitemShut
  {NoStop}%
\bibitem [{\citenamefont {Shen}\ \emph {et~al.}(2012)\citenamefont {Shen},
  \citenamefont {Marcucci}, \citenamefont {Carlson}, \citenamefont {Gandolfi},\
  and\ \citenamefont {Schiavilla}}]{Shen:2012xz}%
  \BibitemOpen
  \bibfield  {author} {\bibinfo {author} {\bibfnamefont {G.}~\bibnamefont
  {Shen}}, \bibinfo {author} {\bibfnamefont {L.~E.}\ \bibnamefont {Marcucci}},
  \bibinfo {author} {\bibfnamefont {J.}~\bibnamefont {Carlson}}, \bibinfo
  {author} {\bibfnamefont {S.}~\bibnamefont {Gandolfi}}, \ and\ \bibinfo
  {author} {\bibfnamefont {R.}~\bibnamefont {Schiavilla}},\ }\href {\doibase
  10.1103/PhysRevC.86.035503} {\bibfield  {journal} {\bibinfo  {journal} {Phys.
  Rev. C}\ }\textbf {\bibinfo {volume} {86}},\ \bibinfo {pages} {035503}
  (\bibinfo {year} {2012})},\ \Eprint {http://arxiv.org/abs/1205.4337}
  {arXiv:1205.4337 [nucl-th]} \BibitemShut {NoStop}%
%%CITATION = ARXIV:1205.4337;%%
\bibitem [{\citenamefont {Golak}\ \emph {et~al.}(2018)\citenamefont {Golak},
  \citenamefont {Skibiński}, \citenamefont {Topolnicki}, \citenamefont
  {Witała}, \citenamefont {Grassi}, \citenamefont {Kamada},\ and\
  \citenamefont {Marcucci}}]{Golak:2018qya}%
  \BibitemOpen
  \bibfield  {author} {\bibinfo {author} {\bibfnamefont {J.}~\bibnamefont
  {Golak}}, \bibinfo {author} {\bibfnamefont {R.}~\bibnamefont {Skibiński}},
  \bibinfo {author} {\bibfnamefont {K.}~\bibnamefont {Topolnicki}}, \bibinfo
  {author} {\bibfnamefont {H.}~\bibnamefont {Witała}}, \bibinfo {author}
  {\bibfnamefont {A.}~\bibnamefont {Grassi}}, \bibinfo {author} {\bibfnamefont
  {H.}~\bibnamefont {Kamada}}, \ and\ \bibinfo {author} {\bibfnamefont {L.~E.}\
  \bibnamefont {Marcucci}},\ }\href {\doibase 10.1103/PhysRevC.98.015501}
  {\bibfield  {journal} {\bibinfo  {journal} {Phys. Rev. C}\ }\textbf {\bibinfo
  {volume} {C8}},\ \bibinfo {pages} {015501} (\bibinfo {year} {2018})},\
  \Eprint {http://arxiv.org/abs/1805.00103} {arXiv:1805.00103 [nucl-th]}
  \BibitemShut {NoStop}%
%%CITATION = ARXIV:1805.00103;%%
\bibitem [{\citenamefont {Carlson}\ and\ \citenamefont
  {Schiavilla}(1992)}]{Carlson:1992ga}%
  \BibitemOpen
  \bibfield  {author} {\bibinfo {author} {\bibfnamefont {J.}~\bibnamefont
  {Carlson}}\ and\ \bibinfo {author} {\bibfnamefont {R.}~\bibnamefont
  {Schiavilla}},\ }\href {\doibase 10.1103/PhysRevLett.68.3682} {\bibfield
  {journal} {\bibinfo  {journal} {Phys. Rev. Lett.}\ }\textbf {\bibinfo
  {volume} {68}},\ \bibinfo {pages} {3682} (\bibinfo {year}
  {1992})}\BibitemShut {NoStop}%
%%CITATION = PRLTA,68,3682;%%
\bibitem [{\citenamefont {{Titterington}}(1985)}]{Titterington:1985}%
  \BibitemOpen
  \bibfield  {author} {\bibinfo {author} {\bibfnamefont {D.~M.}\ \bibnamefont
  {{Titterington}}},\ }\href@noop {} {\bibfield  {journal} {\bibinfo  {journal}
  {Astron. Astrophys.}\ }\textbf {\bibinfo {volume} {144}},\ \bibinfo {pages}
  {381} (\bibinfo {year} {1985})}\BibitemShut {NoStop}%
\bibitem [{\citenamefont {{Gull}}\ and\ \citenamefont
  {{Daniell}}(1978)}]{Gull:1978}%
  \BibitemOpen
  \bibfield  {author} {\bibinfo {author} {\bibfnamefont {S.~F.}\ \bibnamefont
  {{Gull}}}\ and\ \bibinfo {author} {\bibfnamefont {G.~J.}\ \bibnamefont
  {{Daniell}}},\ }\href {\doibase 10.1038/272686a0} {\bibfield  {journal}
  {\bibinfo  {journal} {\nat}\ }\textbf {\bibinfo {volume} {272}},\ \bibinfo
  {pages} {686} (\bibinfo {year} {1978})}\BibitemShut {NoStop}%
\bibitem [{\citenamefont {Skilling}(1989)}]{Skilling:1989}%
  \BibitemOpen
  \bibfield  {author} {\bibinfo {author} {\bibfnamefont {J.}~\bibnamefont
  {Skilling}},\ }\enquote {\bibinfo {title} {Classic maximum entropy},}\ in\
  \href {\doibase 10.1007/978-94-015-7860-8_3} {\emph {\bibinfo {booktitle}
  {Maximum Entropy and Bayesian Method}}},\ \bibinfo {editor} {edited by\
  \bibinfo {editor} {\bibfnamefont {J.}~\bibnamefont {Skilling}}}\ (\bibinfo
  {publisher} {Springer Netherlands},\ \bibinfo {address} {Dordrecht},\
  \bibinfo {year} {1989})\ pp.\ \bibinfo {pages} {45--52}\BibitemShut {NoStop}%
\bibitem [{\citenamefont {{Von Der Linden}}\ \emph {et~al.}(1999)\citenamefont
  {{Von Der Linden}}, \citenamefont {Preuss},\ and\ \citenamefont
  {Dose}}]{VonDerLinden:1999}%
  \BibitemOpen
  \bibfield  {author} {\bibinfo {author} {\bibfnamefont {W.}~\bibnamefont {{Von
  Der Linden}}}, \bibinfo {author} {\bibfnamefont {R.}~\bibnamefont {Preuss}},
  \ and\ \bibinfo {author} {\bibfnamefont {V.}~\bibnamefont {Dose}},\ }in\
  \href@noop {} {\emph {\bibinfo {booktitle} {Maximum Entropy and Bayesian
  Methods}}},\ \bibinfo {editor} {edited by\ \bibinfo {editor} {\bibfnamefont
  {W.}~\bibnamefont {von~der Linden}}, \bibinfo {editor} {\bibfnamefont
  {V.}~\bibnamefont {Dose}}, \bibinfo {editor} {\bibfnamefont {R.}~\bibnamefont
  {Fischer}}, \ and\ \bibinfo {editor} {\bibfnamefont {R.}~\bibnamefont
  {Preuss}}}\ (\bibinfo  {publisher} {Springer Netherlands},\ \bibinfo
  {address} {Dordrecht},\ \bibinfo {year} {1999})\ pp.\ \bibinfo {pages}
  {319--326}\BibitemShut {NoStop}%
\bibitem [{\citenamefont {{Hohenadler}}\ \emph {et~al.}(2005)\citenamefont
  {{Hohenadler}}, \citenamefont {{Neuber}}, \citenamefont {{von der Linden}},
  \citenamefont {{Wellein}}, \citenamefont {{Loos}},\ and\ \citenamefont
  {{Fehske}}}]{Hohenadler:2005}%
  \BibitemOpen
  \bibfield  {author} {\bibinfo {author} {\bibfnamefont {M.}~\bibnamefont
  {{Hohenadler}}}, \bibinfo {author} {\bibfnamefont {D.}~\bibnamefont
  {{Neuber}}}, \bibinfo {author} {\bibfnamefont {W.}~\bibnamefont {{von der
  Linden}}}, \bibinfo {author} {\bibfnamefont {G.}~\bibnamefont {{Wellein}}},
  \bibinfo {author} {\bibfnamefont {J.}~\bibnamefont {{Loos}}}, \ and\ \bibinfo
  {author} {\bibfnamefont {H.}~\bibnamefont {{Fehske}}},\ }\href {\doibase
  10.1103/PhysRevB.71.245111} {\bibfield  {journal} {\bibinfo  {journal}
  {\prb}\ }\textbf {\bibinfo {volume} {71}},\ \bibinfo {eid} {245111} (\bibinfo
  {year} {2005})},\ \Eprint {http://arxiv.org/abs/cond-mat/0412010}
  {arXiv:cond-mat/0412010 [cond-mat.str-el]} \BibitemShut {NoStop}%
\bibitem [{\citenamefont {Donnelly}\ and\ \citenamefont
  {Sick}(1999)}]{Donnelly:1999sw}%
  \BibitemOpen
  \bibfield  {author} {\bibinfo {author} {\bibfnamefont {T.~W.}\ \bibnamefont
  {Donnelly}}\ and\ \bibinfo {author} {\bibfnamefont {I.}~\bibnamefont
  {Sick}},\ }\href {\doibase 10.1103/PhysRevC.60.065502} {\bibfield  {journal}
  {\bibinfo  {journal} {Phys. Rev. C}\ }\textbf {\bibinfo {volume} {60}},\
  \bibinfo {pages} {065502} (\bibinfo {year} {1999})},\ \Eprint
  {http://arxiv.org/abs/nucl-th/9905060} {arXiv:nucl-th/9905060 [nucl-th]}
  \BibitemShut {NoStop}%
%%CITATION = NUCL-TH/9905060;%%
\bibitem [{\citenamefont {Rocco}\ \emph {et~al.}(2017)\citenamefont {Rocco},
  \citenamefont {Alvarez-Ruso}, \citenamefont {Lovato},\ and\ \citenamefont
  {Nieves}}]{Rocco:2017hmh}%
  \BibitemOpen
  \bibfield  {author} {\bibinfo {author} {\bibfnamefont {N.}~\bibnamefont
  {Rocco}}, \bibinfo {author} {\bibfnamefont {L.}~\bibnamefont {Alvarez-Ruso}},
  \bibinfo {author} {\bibfnamefont {A.}~\bibnamefont {Lovato}}, \ and\ \bibinfo
  {author} {\bibfnamefont {J.}~\bibnamefont {Nieves}},\ }\href {\doibase
  10.1103/PhysRevC.96.015504} {\bibfield  {journal} {\bibinfo  {journal} {Phys.
  Rev. C}\ }\textbf {\bibinfo {volume} {96}},\ \bibinfo {pages} {015504}
  (\bibinfo {year} {2017})},\ \Eprint {http://arxiv.org/abs/1701.05151}
  {arXiv:1701.05151 [nucl-th]} \BibitemShut {NoStop}%
%%CITATION = ARXIV:1701.05151;%%
\bibitem [{\citenamefont {Sick}(2001)}]{Sick:2001rh}%
  \BibitemOpen
  \bibfield  {author} {\bibinfo {author} {\bibfnamefont {I.}~\bibnamefont
  {Sick}},\ }\href {\doibase 10.1016/S0146-6410(01)00156-9} {\bibfield
  {journal} {\bibinfo  {journal} {Prog. Part. Nucl. Phys.}\ }\textbf {\bibinfo
  {volume} {47}},\ \bibinfo {pages} {245} (\bibinfo {year} {2001})},\ \Eprint
  {http://arxiv.org/abs/nucl-ex/0208009} {arXiv:nucl-ex/0208009 [nucl-ex]}
  \BibitemShut {NoStop}%
%%CITATION = NUCL-EX/0208009;%%
\bibitem [{\citenamefont {Kingma}\ and\ \citenamefont
  {Ba}(2014)}]{kingma2014adam}%
  \BibitemOpen
  \bibfield  {author} {\bibinfo {author} {\bibfnamefont {D.~P.}\ \bibnamefont
  {Kingma}}\ and\ \bibinfo {author} {\bibfnamefont {J.}~\bibnamefont {Ba}},\
  }\href@noop {} {\bibfield  {journal} {\bibinfo  {journal} {arXiv preprint
  arXiv:1412.6980}\ } (\bibinfo {year} {2014})}\BibitemShut {NoStop}%
\bibitem [{\citenamefont {Abadi}\ \emph {et~al.}(2015)\citenamefont {Abadi},
  \citenamefont {Agarwal}, \citenamefont {Barham}, \citenamefont {Brevdo},
  \citenamefont {Chen}, \citenamefont {Citro}, \citenamefont {Corrado},
  \citenamefont {Davis}, \citenamefont {Dean}, \citenamefont {Devin},
  \citenamefont {Ghemawat}, \citenamefont {Goodfellow}, \citenamefont {Harp},
  \citenamefont {Irving}, \citenamefont {Isard}, \citenamefont {Jia},
  \citenamefont {Jozefowicz}, \citenamefont {Kaiser}, \citenamefont {Kudlur},
  \citenamefont {Levenberg}, \citenamefont {Man\'{e}}, \citenamefont {Monga},
  \citenamefont {Moore}, \citenamefont {Murray}, \citenamefont {Olah},
  \citenamefont {Schuster}, \citenamefont {Shlens}, \citenamefont {Steiner},
  \citenamefont {Sutskever}, \citenamefont {Talwar}, \citenamefont {Tucker},
  \citenamefont {Vanhoucke}, \citenamefont {Vasudevan}, \citenamefont
  {Vi\'{e}gas}, \citenamefont {Vinyals}, \citenamefont {Warden}, \citenamefont
  {Wattenberg}, \citenamefont {Wicke}, \citenamefont {Yu},\ and\ \citenamefont
  {Zheng}}]{tensorflow2015-whitepaper}%
  \BibitemOpen
  \bibfield  {author} {\bibinfo {author} {\bibfnamefont {M.}~\bibnamefont
  {Abadi}}, \bibinfo {author} {\bibfnamefont {A.}~\bibnamefont {Agarwal}},
  \bibinfo {author} {\bibfnamefont {P.}~\bibnamefont {Barham}}, \bibinfo
  {author} {\bibfnamefont {E.}~\bibnamefont {Brevdo}}, \bibinfo {author}
  {\bibfnamefont {Z.}~\bibnamefont {Chen}}, \bibinfo {author} {\bibfnamefont
  {C.}~\bibnamefont {Citro}}, \bibinfo {author} {\bibfnamefont {G.~S.}\
  \bibnamefont {Corrado}}, \bibinfo {author} {\bibfnamefont {A.}~\bibnamefont
  {Davis}}, \bibinfo {author} {\bibfnamefont {J.}~\bibnamefont {Dean}},
  \bibinfo {author} {\bibfnamefont {M.}~\bibnamefont {Devin}}, \bibinfo
  {author} {\bibfnamefont {S.}~\bibnamefont {Ghemawat}}, \bibinfo {author}
  {\bibfnamefont {I.}~\bibnamefont {Goodfellow}}, \bibinfo {author}
  {\bibfnamefont {A.}~\bibnamefont {Harp}}, \bibinfo {author} {\bibfnamefont
  {G.}~\bibnamefont {Irving}}, \bibinfo {author} {\bibfnamefont
  {M.}~\bibnamefont {Isard}}, \bibinfo {author} {\bibfnamefont
  {Y.}~\bibnamefont {Jia}}, \bibinfo {author} {\bibfnamefont {R.}~\bibnamefont
  {Jozefowicz}}, \bibinfo {author} {\bibfnamefont {L.}~\bibnamefont {Kaiser}},
  \bibinfo {author} {\bibfnamefont {M.}~\bibnamefont {Kudlur}}, \bibinfo
  {author} {\bibfnamefont {J.}~\bibnamefont {Levenberg}}, \bibinfo {author}
  {\bibfnamefont {D.}~\bibnamefont {Man\'{e}}}, \bibinfo {author}
  {\bibfnamefont {R.}~\bibnamefont {Monga}}, \bibinfo {author} {\bibfnamefont
  {S.}~\bibnamefont {Moore}}, \bibinfo {author} {\bibfnamefont
  {D.}~\bibnamefont {Murray}}, \bibinfo {author} {\bibfnamefont
  {C.}~\bibnamefont {Olah}}, \bibinfo {author} {\bibfnamefont {M.}~\bibnamefont
  {Schuster}}, \bibinfo {author} {\bibfnamefont {J.}~\bibnamefont {Shlens}},
  \bibinfo {author} {\bibfnamefont {B.}~\bibnamefont {Steiner}}, \bibinfo
  {author} {\bibfnamefont {I.}~\bibnamefont {Sutskever}}, \bibinfo {author}
  {\bibfnamefont {K.}~\bibnamefont {Talwar}}, \bibinfo {author} {\bibfnamefont
  {P.}~\bibnamefont {Tucker}}, \bibinfo {author} {\bibfnamefont
  {V.}~\bibnamefont {Vanhoucke}}, \bibinfo {author} {\bibfnamefont
  {V.}~\bibnamefont {Vasudevan}}, \bibinfo {author} {\bibfnamefont
  {F.}~\bibnamefont {Vi\'{e}gas}}, \bibinfo {author} {\bibfnamefont
  {O.}~\bibnamefont {Vinyals}}, \bibinfo {author} {\bibfnamefont
  {P.}~\bibnamefont {Warden}}, \bibinfo {author} {\bibfnamefont
  {M.}~\bibnamefont {Wattenberg}}, \bibinfo {author} {\bibfnamefont
  {M.}~\bibnamefont {Wicke}}, \bibinfo {author} {\bibfnamefont
  {Y.}~\bibnamefont {Yu}}, \ and\ \bibinfo {author} {\bibfnamefont
  {X.}~\bibnamefont {Zheng}},\ }\href {http://tensorflow.org/} {\enquote
  {\bibinfo {title} {{TensorFlow}: Large-scale machine learning on
  heterogeneous systems},}\ } (\bibinfo {year} {2015}),\ \bibinfo {note}
  {software available from tensorflow.org}\BibitemShut {NoStop}%
\bibitem [{\citenamefont {Birge}(1932)}]{Birge:1932}%
  \BibitemOpen
  \bibfield  {author} {\bibinfo {author} {\bibfnamefont {R.~T.}\ \bibnamefont
  {Birge}},\ }\href {\doibase 10.1103/PhysRev.40.207} {\bibfield  {journal}
  {\bibinfo  {journal} {Phys. Rev.}\ }\textbf {\bibinfo {volume} {40}},\
  \bibinfo {pages} {207} (\bibinfo {year} {1932})}\BibitemShut {NoStop}%
\bibitem [{\citenamefont {Efros}\ \emph {et~al.}(1994)\citenamefont {Efros},
  \citenamefont {Leidemann},\ and\ \citenamefont {Orlandini}}]{Efros:1994iq}%
  \BibitemOpen
  \bibfield  {author} {\bibinfo {author} {\bibfnamefont {V.~D.}\ \bibnamefont
  {Efros}}, \bibinfo {author} {\bibfnamefont {W.}~\bibnamefont {Leidemann}}, \
  and\ \bibinfo {author} {\bibfnamefont {G.}~\bibnamefont {Orlandini}},\ }\href
  {\doibase 10.1016/0370-2693(94)91355-2} {\bibfield  {journal} {\bibinfo
  {journal} {Phys. Lett. B}\ }\textbf {\bibinfo {volume} {338}},\ \bibinfo
  {pages} {130} (\bibinfo {year} {1994})},\ \Eprint
  {http://arxiv.org/abs/nucl-th/9409004} {arXiv:nucl-th/9409004} \BibitemShut
  {NoStop}%
\bibitem [{\citenamefont {Efros}\ \emph {et~al.}(1997)\citenamefont {Efros},
  \citenamefont {Leidemann},\ and\ \citenamefont {Orlandini}}]{Efros:1997jf}%
  \BibitemOpen
  \bibfield  {author} {\bibinfo {author} {\bibfnamefont {V.~D.}\ \bibnamefont
  {Efros}}, \bibinfo {author} {\bibfnamefont {W.}~\bibnamefont {Leidemann}}, \
  and\ \bibinfo {author} {\bibfnamefont {G.}~\bibnamefont {Orlandini}},\ }\href
  {\doibase 10.1103/PhysRevLett.78.432} {\bibfield  {journal} {\bibinfo
  {journal} {Phys. Rev. Lett.}\ }\textbf {\bibinfo {volume} {78}},\ \bibinfo
  {pages} {432} (\bibinfo {year} {1997})},\ \Eprint
  {http://arxiv.org/abs/nucl-th/9701004} {arXiv:nucl-th/9701004} \BibitemShut
  {NoStop}%
\bibitem [{\citenamefont {Bacca}\ \emph {et~al.}(2002)\citenamefont {Bacca},
  \citenamefont {Marchisio}, \citenamefont {Barnea}, \citenamefont
  {Leidemann},\ and\ \citenamefont {Orlandini}}]{Bacca:2001kr}%
  \BibitemOpen
  \bibfield  {author} {\bibinfo {author} {\bibfnamefont {S.}~\bibnamefont
  {Bacca}}, \bibinfo {author} {\bibfnamefont {M.~A.}\ \bibnamefont
  {Marchisio}}, \bibinfo {author} {\bibfnamefont {N.}~\bibnamefont {Barnea}},
  \bibinfo {author} {\bibfnamefont {W.}~\bibnamefont {Leidemann}}, \ and\
  \bibinfo {author} {\bibfnamefont {G.}~\bibnamefont {Orlandini}},\ }\href
  {\doibase 10.1103/PhysRevLett.89.052502} {\bibfield  {journal} {\bibinfo
  {journal} {Phys. Rev. Lett.}\ }\textbf {\bibinfo {volume} {89}},\ \bibinfo
  {pages} {052502} (\bibinfo {year} {2002})},\ \Eprint
  {http://arxiv.org/abs/nucl-th/0112067} {arXiv:nucl-th/0112067} \BibitemShut
  {NoStop}%
\bibitem [{\citenamefont {Bacca}\ \emph {et~al.}(2009)\citenamefont {Bacca},
  \citenamefont {Barnea}, \citenamefont {Leidemann},\ and\ \citenamefont
  {Orlandini}}]{Bacca:2008tb}%
  \BibitemOpen
  \bibfield  {author} {\bibinfo {author} {\bibfnamefont {S.}~\bibnamefont
  {Bacca}}, \bibinfo {author} {\bibfnamefont {N.}~\bibnamefont {Barnea}},
  \bibinfo {author} {\bibfnamefont {W.}~\bibnamefont {Leidemann}}, \ and\
  \bibinfo {author} {\bibfnamefont {G.}~\bibnamefont {Orlandini}},\ }\href
  {\doibase 10.1103/PhysRevLett.102.162501} {\bibfield  {journal} {\bibinfo
  {journal} {Phys. Rev. Lett.}\ }\textbf {\bibinfo {volume} {102}},\ \bibinfo
  {pages} {162501} (\bibinfo {year} {2009})},\ \Eprint
  {http://arxiv.org/abs/0811.4624} {arXiv:0811.4624 [nucl-th]} \BibitemShut
  {NoStop}%
\bibitem [{\citenamefont {Bacca}\ \emph {et~al.}(2013)\citenamefont {Bacca},
  \citenamefont {Barnea}, \citenamefont {Hagen}, \citenamefont {Orlandini},\
  and\ \citenamefont {Papenbrock}}]{Bacca:2013dma}%
  \BibitemOpen
  \bibfield  {author} {\bibinfo {author} {\bibfnamefont {S.}~\bibnamefont
  {Bacca}}, \bibinfo {author} {\bibfnamefont {N.}~\bibnamefont {Barnea}},
  \bibinfo {author} {\bibfnamefont {G.}~\bibnamefont {Hagen}}, \bibinfo
  {author} {\bibfnamefont {G.}~\bibnamefont {Orlandini}}, \ and\ \bibinfo
  {author} {\bibfnamefont {T.}~\bibnamefont {Papenbrock}},\ }\href {\doibase
  10.1103/PhysRevLett.111.122502} {\bibfield  {journal} {\bibinfo  {journal}
  {Phys. Rev. Lett.}\ }\textbf {\bibinfo {volume} {111}},\ \bibinfo {pages}
  {122502} (\bibinfo {year} {2013})},\ \Eprint {http://arxiv.org/abs/1303.7446}
  {arXiv:1303.7446 [nucl-th]} \BibitemShut {NoStop}%
\bibitem [{\citenamefont {Birkhan}\ \emph {et~al.}(2017)\citenamefont {Birkhan}
  \emph {et~al.}}]{Birkhan:2016qkr}%
  \BibitemOpen
  \bibfield  {author} {\bibinfo {author} {\bibfnamefont {J.}~\bibnamefont
  {Birkhan}} \emph {et~al.},\ }\href {\doibase 10.1103/PhysRevLett.118.252501}
  {\bibfield  {journal} {\bibinfo  {journal} {Phys. Rev. Lett.}\ }\textbf
  {\bibinfo {volume} {118}},\ \bibinfo {pages} {252501} (\bibinfo {year}
  {2017})},\ \Eprint {http://arxiv.org/abs/1611.07072} {arXiv:1611.07072
  [nucl-ex]} \BibitemShut {NoStop}%
\bibitem [{\citenamefont {Simonis}\ \emph {et~al.}(2019)\citenamefont
  {Simonis}, \citenamefont {Bacca},\ and\ \citenamefont
  {Hagen}}]{Simonis:2019spj}%
  \BibitemOpen
  \bibfield  {author} {\bibinfo {author} {\bibfnamefont {J.}~\bibnamefont
  {Simonis}}, \bibinfo {author} {\bibfnamefont {S.}~\bibnamefont {Bacca}}, \
  and\ \bibinfo {author} {\bibfnamefont {G.}~\bibnamefont {Hagen}},\ }\href
  {\doibase 10.1140/epja/i2019-12825-0} {\bibfield  {journal} {\bibinfo
  {journal} {Eur. Phys. J. A}\ }\textbf {\bibinfo {volume} {55}},\ \bibinfo
  {pages} {241} (\bibinfo {year} {2019})},\ \Eprint
  {http://arxiv.org/abs/1905.02055} {arXiv:1905.02055 [nucl-th]} \BibitemShut
  {NoStop}%
\bibitem [{\citenamefont {Roggero}(2020)}]{Roggero:2020qoz}%
  \BibitemOpen
  \bibfield  {author} {\bibinfo {author} {\bibfnamefont {A.}~\bibnamefont
  {Roggero}},\ }\href {\doibase 10.1103/PhysRevA.102.022409} {\bibfield
  {journal} {\bibinfo  {journal} {Phys. Rev. A}\ }\textbf {\bibinfo {volume}
  {102}},\ \bibinfo {pages} {022409} (\bibinfo {year} {2020})},\ \Eprint
  {http://arxiv.org/abs/2004.04889} {arXiv:2004.04889 [quant-ph]} \BibitemShut
  {NoStop}%
\end{thebibliography}%

\end{document}